\documentclass[a4paper]{spie}  

\usepackage{graphicx}
\usepackage{txfonts}
\usepackage[colorlinks=true,citecolor=blue]{hyperref}
\usepackage{xspace}
\usepackage{array}
\usepackage{bm}
\usepackage{gensymb}

\newcommand{\as}{\hbox{$^{\prime\prime}$}\xspace}

\newcommand{\cpup}{c/p\xspace}

\newcommand{\corgisim}{\texttt{corgisim}\xspace}
\newcommand{\dmconf}{\ensuremath{5 \times 10^{-9}}\xspace}

\title{The Roman Coronagraph Community Participation Program:\\Using the Zernike wavefront sensor for a full characterisation of the Roman Space Telescope and the Coronagraph Instrument}

\author[a]{Arthur Vigan}
\author[b]{Alexis Bidot}
\author[a]{Vincent Chambouleyron}
\author[d]{Garreth Ruane}
\author[b]{Laurent Pueyo}
\author[c]{Mamadou N'Diaye}
\author[a]{Kjetil Dohlen}

\affil[a]{Aix Marseille Univ, CNRS, CNES, LAM, Marseille, France}
\affil[b]{Space Telescope Science Institute, 3700 San Martin Drive, Baltimore, MD 21218, USA}
\affil[c]{Laboratoire Lagrange, Université Côte d'Azur, Observatoire de la Côte d'Azur, CNRS, Nice, France}
\affil[d]{Jet Propulsion Laboratory, California Institute of Technology}

\authorinfo{Further author information: \href{mailto:arthur.vigan@lam.fr}{arthur.vigan@lam.fr}}

\begin{document}
\maketitle

\begin{abstract}
The Roman Space Telescope Coronagraph Instrument (CGI) will demonstrate a series of technologies and techniques to enable the direct detection of reflected-light planets with space-based observatories. Among the several available observing modes and coronagraphic devices embarked in CGI, there is the transmissive dual-path Zernike wavefront sensor (ZWFS) that could be used to directly measure optical aberrations in the system. The dual-path ZWFS is currently unsupported, but in this work we advocate for the commissioning of this unique observing mode. We investigate the sensitivity of the ZWFS using CGI simulator and other tools developed and supported by the Roman community participation program (CPP). This type of analysis is crucial for understanding the stability of the Roman observatory and to prepare the path towards HWO.
\end{abstract}


\section{Introduction}
\label{sec:introduction}

One of the major endeavours of the exoplanet field in the coming decades is the direct detection of Earth analogues using either nulling interferometry or high-contrast imaging (HCI). For imaging, the Habitable Worlds Observatory (HWO) holds great promises with a potentially large diameter and an off-axis design, which would provide an excellent setup for extremely efficient starlight suppression with coronagraphy and other wavefront control techniques. The technical requirements for a coronagraphic instrument on HWO are driven by the goal of imaging telluric planets in the habitable zone of solar-type stars, which translates into contrasts of the order of $10^{10}$ and angular separations of the order of 0.1\as in the visible. To reach this level of performance, the coronagraphic instrument must create a region of extremely high contrast in the vicinity of the start using wavefront control, usually referred to as a ``dark hole'' (DH).

These techniques are now well established and have been demonstrated in the laboratory\cite{Potier2020} and on-sky\cite{Potier2022} with ground-based facilities. They usually rely on focal-plane wavefront sensing and electric field conjugation (EFC), which are used iteratively to shape the wavefront using deformable mirrors (DMs) and dig the DH\cite{Give'on2007,Pueyo2009,Give'on2011}. Depending on the accuracy of the instrument model used, the number of iterations necessary to reach a given level of contrast can increase significantly, which has an impact on science operations. To reduce the number of necessary iterations, the DH digging will be done on a bright reference star (typically with $V < 2$) before slewing the observatory to the science target. The Roman Coronagraph\cite{Bailey2023} will be the first to use these techniques in space\cite{Cady2025} in preparation for HWO.

The level of residual speckles and their decorrelation in the DH region is directly related to precision at which the wavefront can be controlled and stabilised over time\cite{Soummer2007}. For telluric planets at contrasts of the order of $10^{10}$, wavefront errors that can be tolerated at mid-spatial frequencies (3-20\,cycle/pupil or \cpup) are of the order of a few picometers (pm) over long periods of time (tens to hundreds of hours)\cite{Juanola-Parramon2019,LUVOIR2019}. Even at very stable locations like the second Lagrangian point (L2), maintaining passively this level of stability is next to impossible. Slewing the observatory from the reference star to the science target, or rolling it to induce angular diversity into the data during observations, are enough to change the solar insulation, which will propagate to variations of the wavefront and of the residual speckles in the DH. And even without moving the observatory, the wavefront will always slowly vary over long periods of observing time, requiring some adjustments to the DMs shape to maintain the DH\cite{Redmond2024}. Switching back to a bright reference star to do a DH touch up is inefficient and costly, which is why this operation should be kept as a last resort.

A more efficient approach is to sense the wavefront variations in parallel to the science observations, using light at a different wavelength from the science reflected back from the coronagraphic focal-plane mask (FPM). The Zernike wavefront sensor (ZWFS), based on Zernike's phase contrast method\cite{Zernike1934}, is usually considered the best WFS for this type of applications due to use simplicity, sensitivity and spatial resolution\cite{N'Diaye2013,N'Diaye2016,Ruane2020,Vigan2019}. Moreover, it can relatively easily be implemented directly on the FPM as a phase-shifting dimple combined with a dichroic function. This is the approached implemented in the Roman coronagraph (see Sect.~\ref{sec:roman_zwfs}) and considered for HWO's precursor studies (HabEx\cite{Gaudi2020} and LUVOIR\cite{LUVOIR2019}). An in situ ZWFS also offers important benefits in terms of diagnostics of the observatory or coronagraph instrument\cite{Sauvage2015,Vigan2022}.

In this paper, we investigate the use of the dual-path ZWFS implemented in the Roman Coronagraph for a full characterisation of the observatory. This study is performed using the tools developed in the context of the Community Participation Program (CPP). In Sect.~\ref{sec:roman_zwfs} we present the different flavours of ZWFS implemented in Roman, and in particular the dual-path ZWFS that offers the best spatial sampling of the pupil. Then, in Sect.~\ref{sec:perf} we assess the expected performance of the dual-path ZWFS in different configurations, and in Sect.~\ref{sec:os11} we simulate a full sequence based on the OS11 observing scenario. Finally, in Sect.~\ref{sec:conclusions} we present some perspectives in the context of the development of HWO.

\section{The Roman Coronagraph ZWFS}
\label{sec:roman_zwfs}

The Roman Coronagraph includes several ZWFS\cite{Riggs2025}. The main one is at the heart of the low-order wavefront sensor (LOWFS), which is part of the baseline for most coronagraphic observations\cite{Seo2025}. The LOWFS is tasked with maintaining the optical wavefront stability requirement at low spatial frequencies, which are essential to maintain the performance of the coronagraph, in particular the hybrid Lyot coronagraph. The LOWFS operates in parallel with the scientific observations and is able to sense low-order Zernike terms from Z2 to Z11 (Noll Zernike order). It works by producing a pupil image on LOCAM (the LOWFS camera) with a sampling of 38 pixels across the pupil. One of the particularities of the LOWFS is that it operates on the starlight rejected by the reflective FPM, which means that it is a spatially filtered ZWFS. The LOWFS uses modal sensing for pointing control to reduce the impact of noise. It corrects tip-tilt variations using a fast steering mechanism (FSM) with a control sampling rate of 1\,kHz and a bandwidth of 20\,Hz. And higher orders from Z4 to Z11 are compensated using DM1 with a sampling rate of 0.1\,Hz and a bandwidth of $1.6 \times 10^{-3}$\,Hz (i.e., $\sim$10\,min).

Because of its spatially-filtered design and of the limited sampling of the pupil on LOCAM, the sensitivity of LOWFS to higher order Zernike terms is limited. In particular, it is not able to sample correctly individual actuators of the DMs. This is why several higher-order ZWFS have also been included in the Coronagraph in the focal plane (mask) alignment mechanism (FPAM). They are, however, untested and unsupported at the time of launch due to the limited resources from the instrument team to fully support additional observing modes and the small amount of time available during TVAC. A family of these additional ZWFS are fully reflective masks included in the SPC12 and SPC34 substrates of the FPAM, which carry the FPMs for the shaped-pupil coronagraph (SPC) for band 1/2 and 3/4, respectively. These masks send all the incoming light towards LOCAM, so their WFS capabilities are only limited by the sampling of the pupil on LOCAM, that is approximately 25\,\cpup.

Finally, a second family of these masks are so-called ``dual-path'' ZWFS because they are designed to operate both in reflection on LOCAM and in transmission on EXCAM (the science camera of the Coronagraph). The family is composed of seven masks with different dimple depths, from 174 to 244\,nm, which are engraved in the polydimethylglutarimide (PMGI) material used to encode the phase part of the HLC focal-plane masks\cite{Riggs2025}. The ZWFS produce a pupil image of almost 300 pixels in diameter, potentially enabling phase measurements up to 150\,\cpup when doing pixel-wise sensing. In addition to its sensitivity to high spatial frequencies, the dual-path ZWFS also offers the advantage that is enables LOWFS operations in parallel of the measurements on EXCAM thanks to the reflective part. This is an important aspect to maintain the pointing control loop of LOWFS and to avoid losing in sensitivity due to the point spread function (PSF) jitter in the focal-plane when LOWFS is not in operation (of the order of 9\,mas\,rms). One caveat of this sensor is that the reflected parts\footnote{But much less for the transmitted part.} of the beam are extremely chromatic due to the way the masks were manufactured: thin-film effects due to the layer of PMGI on top of an AR-coated fused silica create strong attenuations in the reflected part, which could have an impact on the operations of the LOWFS. Estimates done at JPL show that it should work but some tuning of the LOWFS gain may be needed (private communication). In the present paper we will focus on the expected performance of the dual-path ZWFS.

\section{Performance assessment}
\label{sec:perf}

\subsection{Context}

The performance of the dual-path ZWFS was assessed using simulations performed with \href{https://github.com/roman-corgi/corgisim}{\corgisim}, a simulation suite for the Roman Space Telescope Coronagraph developed by the Community Participation Program (CPP) team. It is based on the official diffraction model of the instrument\cite{Krist2023} simulated using the PROPER library\cite{Krist2007}. \corgisim offers the possibility to simulate data in all the modes potentially supported by the coronagraph, including the EMCCD effects. For ZWFS measurements the simulation is performed with the pupil lens inserted into the beam to be able to image the instrument pupil. There is also the possibility to apply custom voltages on the DMs, with predefined states corresponding to a flat configuration and after digging a DH up to a certain contrast level ($3\times10^{-8}$, \dmconf, and $2\times10^{-9}$).

For this study, we use the ZWFS with an absolute phase reconstruction using the \href{https://github.com/avigan/pyZELDA}{\texttt{pyZELDA}} python package\cite{pyZELDA}, which provides as output optical path difference (OPD) maps calibrated in nanometers. The reconstruction is based on a second order polynomial reconstruction\cite{N'Diaye2013,N'Diaye2016}, which is commonly used for its simplicity, although there are alternatives providing a better reconstruction especially when phase errors are outside the linear approximation\cite{Chambouleyron2024}. The reconstruction is based on two images: one obtained with the ZWFS mask inserted into the beam, and one obtained without the mask (or with the mask shifted so that the core of the PSF does not hit the phase-shifting dimple), often referred to as the clear pupil image, which is used for normalisation.

\begin{figure}
  \centering
  \includegraphics[height=5cm]{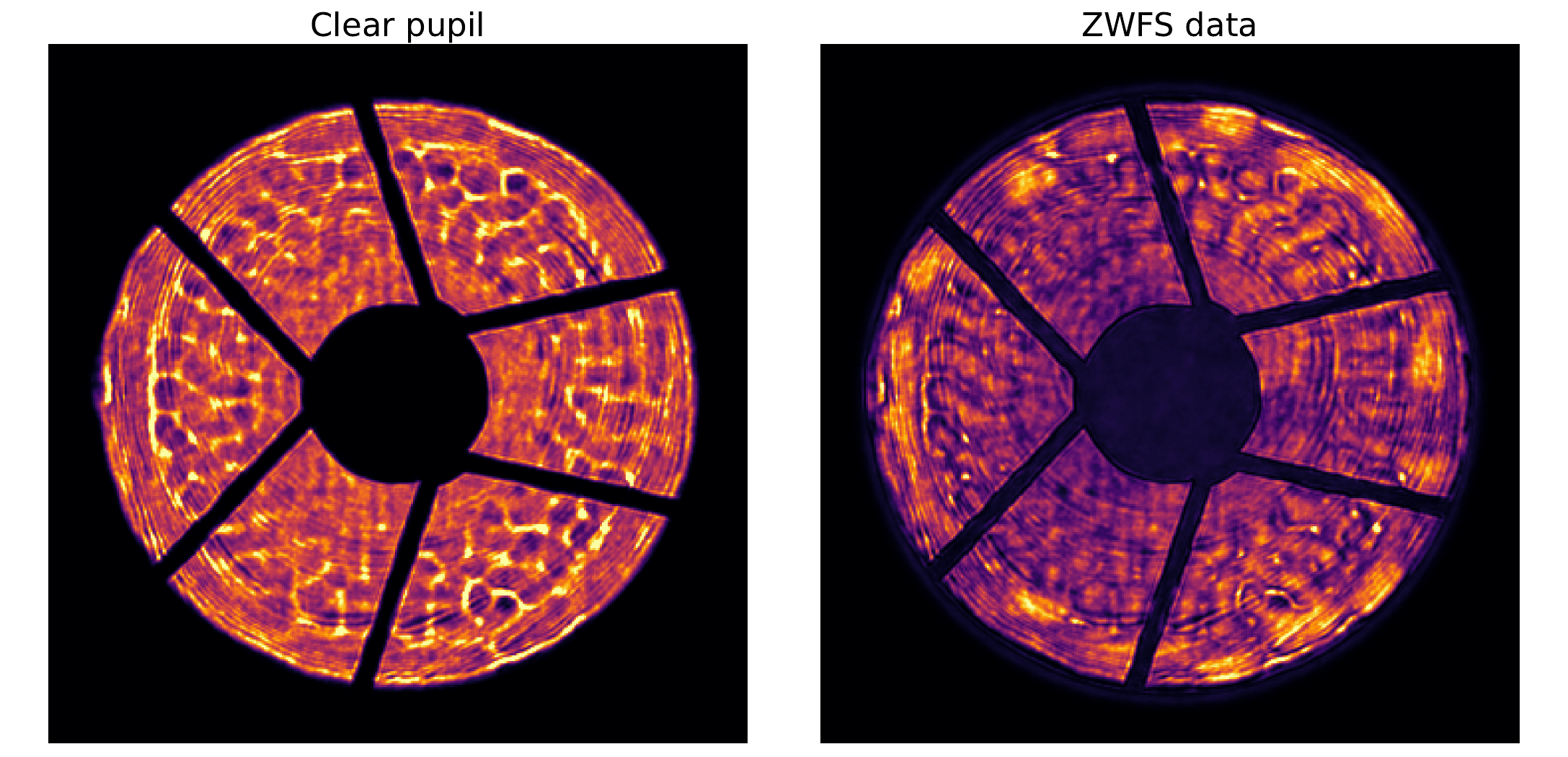}
  \includegraphics[height=5cm]{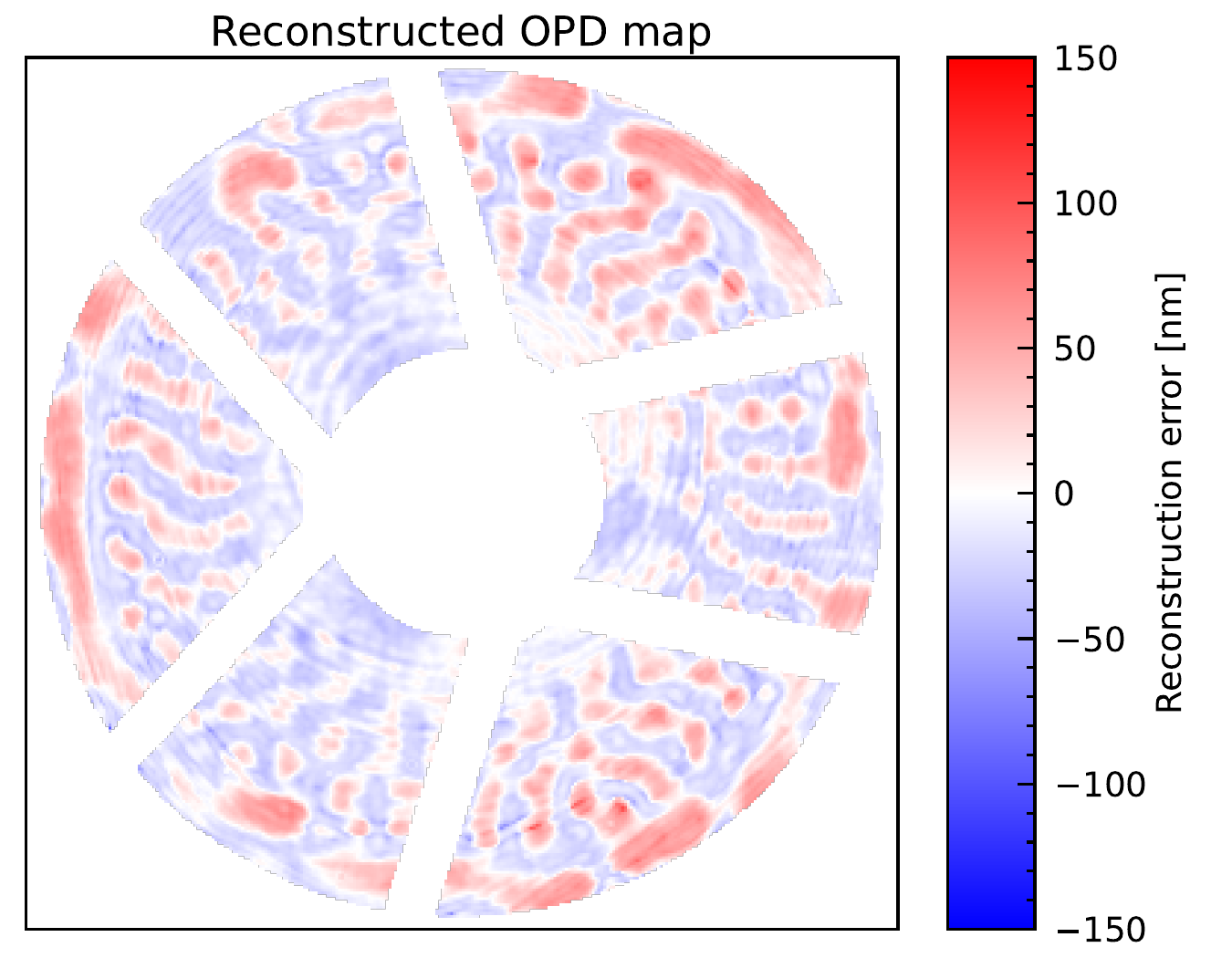}
  \caption{Example of noiseless ZWFS images obtained in the Band 1 filter with the \dmconf DMs configuration. The left image corresponds to the clear pupil, while the centre one is the ZWFS measurement. The right image corresponds to the reconstructed OPD map calibrated in nm. In the reconstruction, the pupil footprint has been slightly enlarged to mask edge effects creating spikes in the reconstruction. The strong intensity and OPD variations visible in the images are the effect of the DH digging process, which applies very specific shapes on the DMs to create the DH in focal-plane images.}
  \label{fig:zwfs_images}
\end{figure}

One of the main problem in HCI is the slow evolution of aberrations, which will induce a decorrelation of speckles in the focal plane. The processing of Roman Coronagraph data will include both reference differential imaging (RDI) and roll subtractions to remove residual speckles in science images. In this context, the decorrelation of speckles with time is an important limitation, which is why typical Roman observing sequences will include DH touch-up in long sequences to make sure that the speckle noise floor remains at a specified level\cite{Krist2023}. The presence of the LOWFS is a mitigation strategy to correct for the evolution of low-order aberrations, which usually are the most limiting for the performance of the coronagraph. However, it would still be interesting to be able to monitor how higher order spatial frequencies evolve with time. This can provide a wealth of information on both the observatory and the Coronagraph.

The dual-path ZWFS is an ideal tool for these kinds of applications. The evolution of aberrations with a ZWFS has already demonstrated its power for ground-based instrumentation using differential measurements\cite{Vigan2022}, but the Roman Coronagraph is really the first time that we can investigate these issues in space. This is highly relevant for HWO, for which Roman is a precursor. There are, however, potential limitations. An important one is that the dual-path ZWFS has not been tested in TVAC and is currently not supported for in-space operations. While data acquisition should be fairly straightforward in this observing mode, there is guarantee that 1/\,the stellar PSF can be accurately centred on the dimple, and 2/\,the LOWFS can actually close the loop on the reflected part of the PSF (see Sect.~\ref{sec:roman_zwfs}).

To assess if the dual-path ZWFS in Roman could be used to monitor the evolution of aberrations, even considering the potential caveats stated above, we perform different sets of simulations. In particular, we study the impact of a PSF offset on the ZWFS mask (Sect.~\ref{sec:perf:offset}), the effect of jitter (Sect.~\ref{sec:perf:jitter}) and the sensitivity of the sensor to very small phase variations (Sect.~\ref{sec:perf:small_errors}).

\subsection{Offset on the mask}
\label{sec:perf:offset}

\begin{figure}
  \centering
  \includegraphics[width=0.49\textwidth]{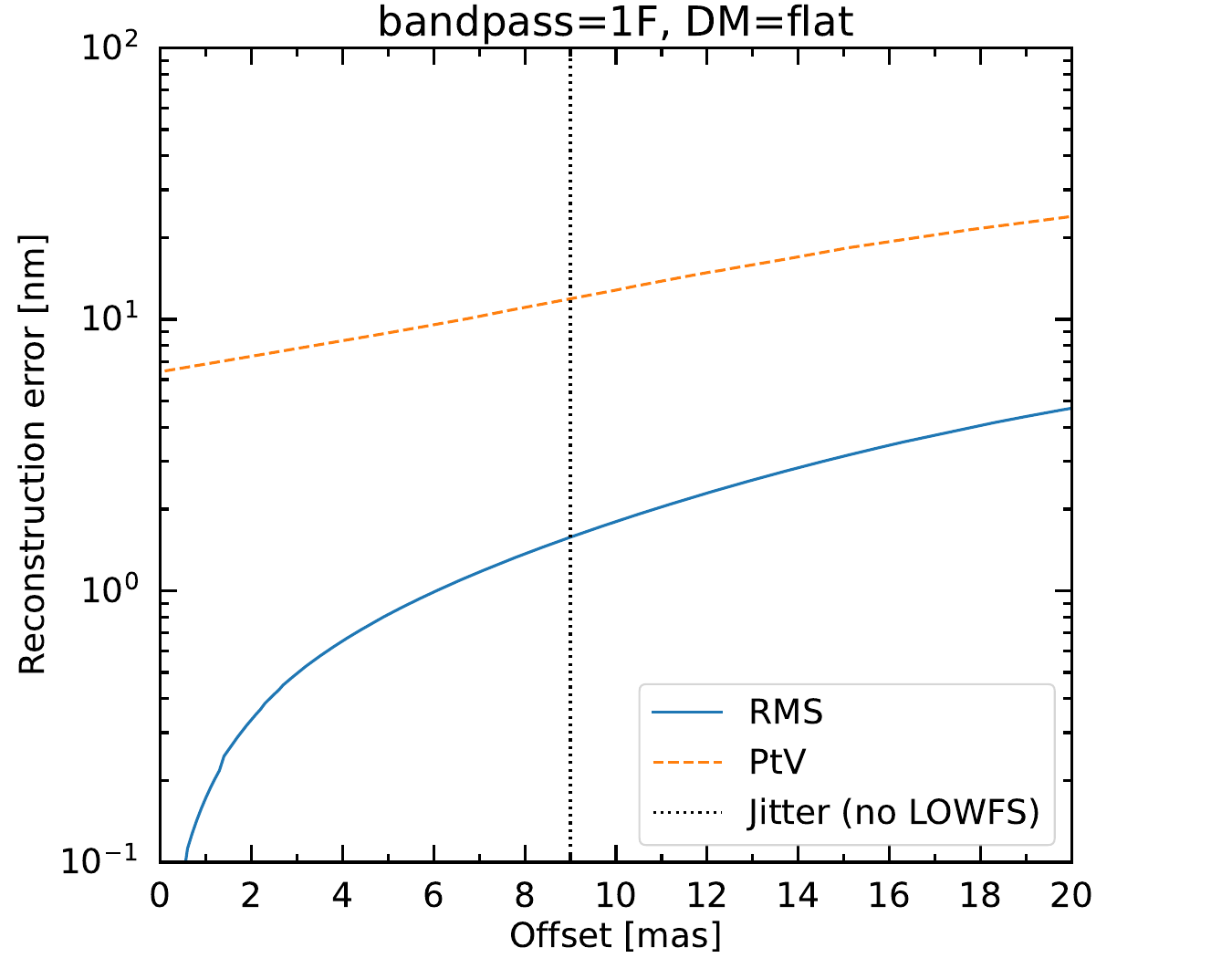}
  \includegraphics[width=0.49\textwidth]{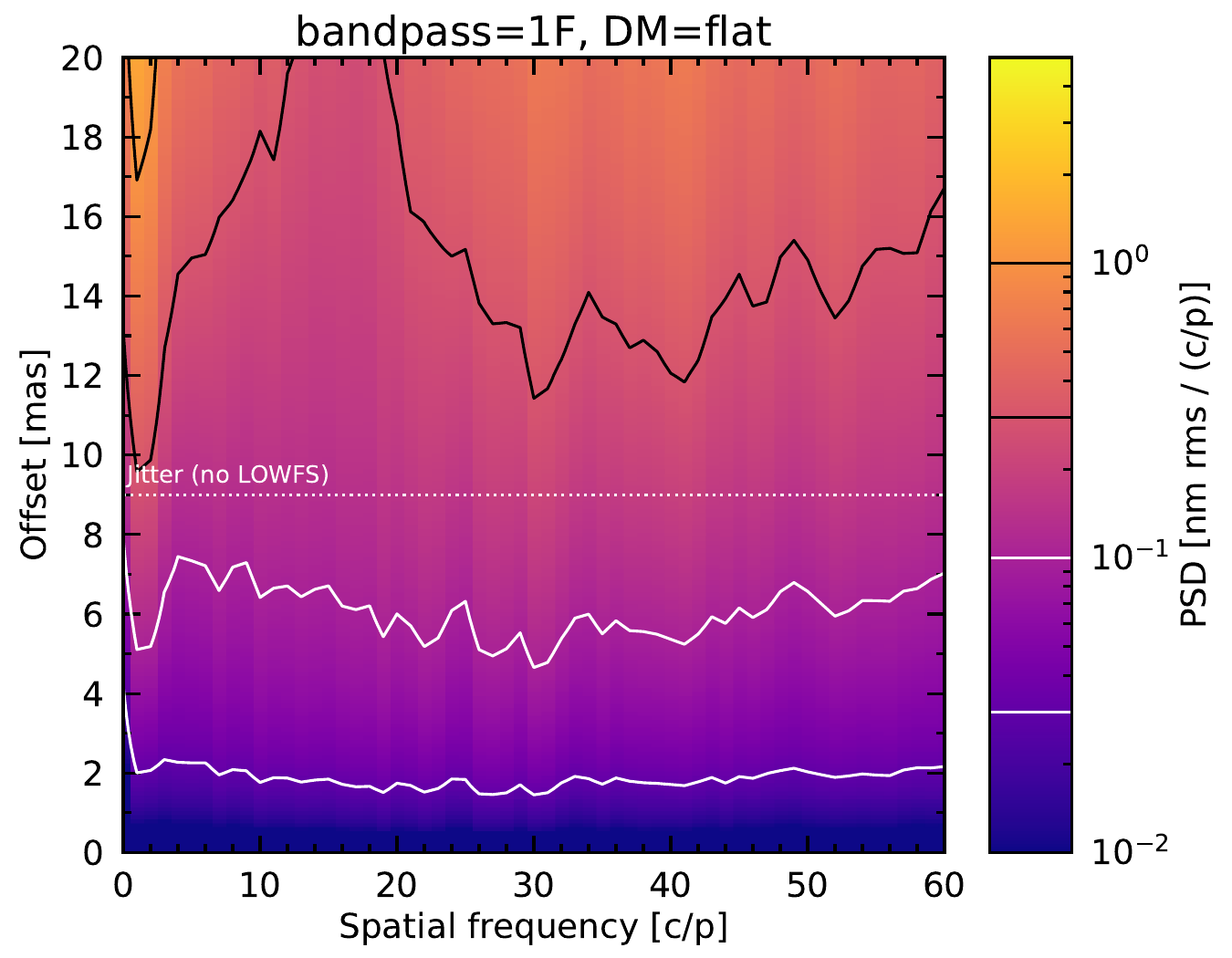}
  \caption{Wavefront reconstruction error with the ZWFS in case of an offset of the stellar PSF on the mask, assuming a flat DMs configuration. The left plot shows the full wavefront reconstruction error (in nm rm), while the right plot shows the decomposition in terms of spatial frequency (in nm rms / (\cpup)).}
  \label{fig:offset_dm=flat}
\end{figure}

\begin{figure}
  \centering
  \includegraphics[width=0.49\textwidth]{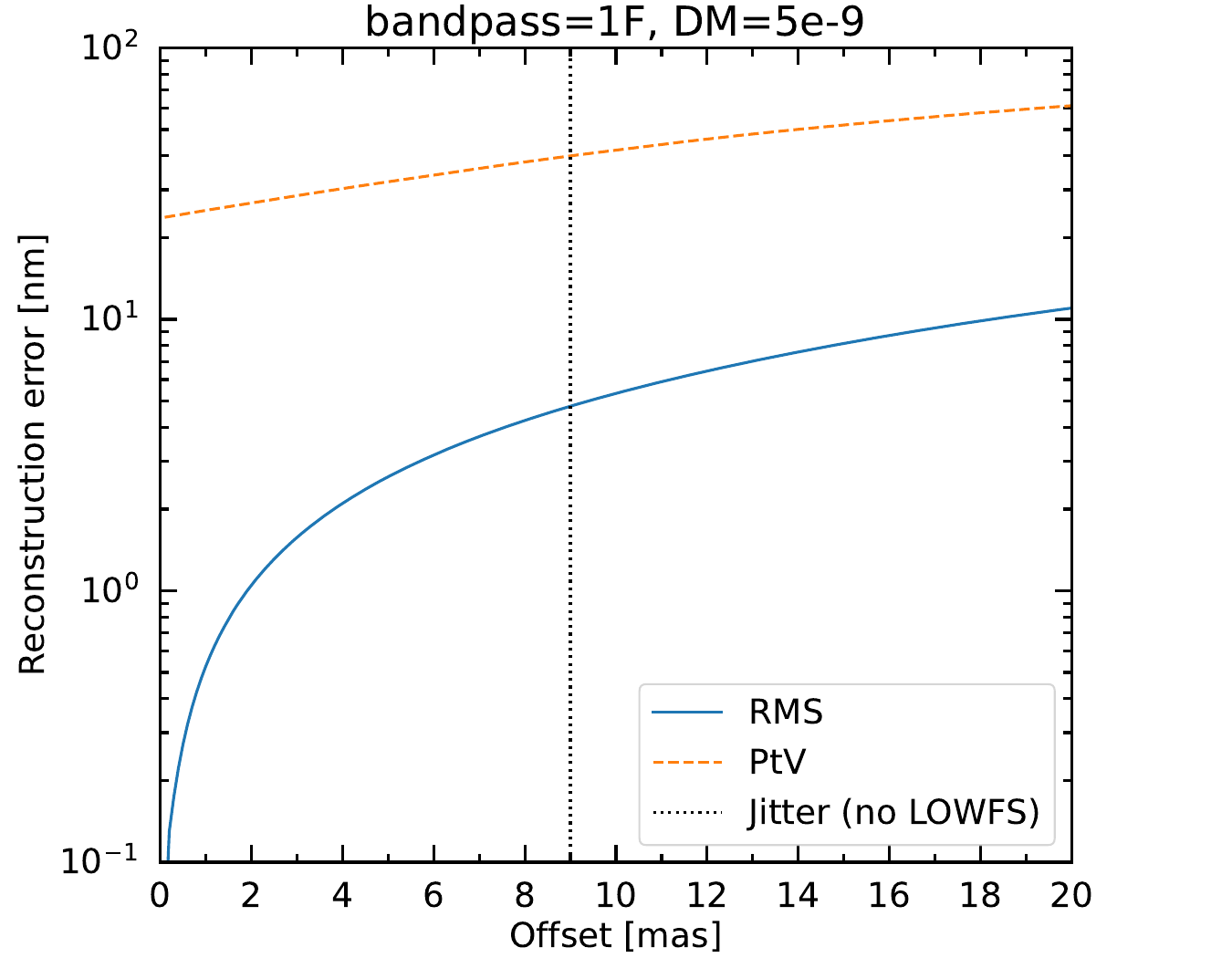}
  \includegraphics[width=0.49\textwidth]{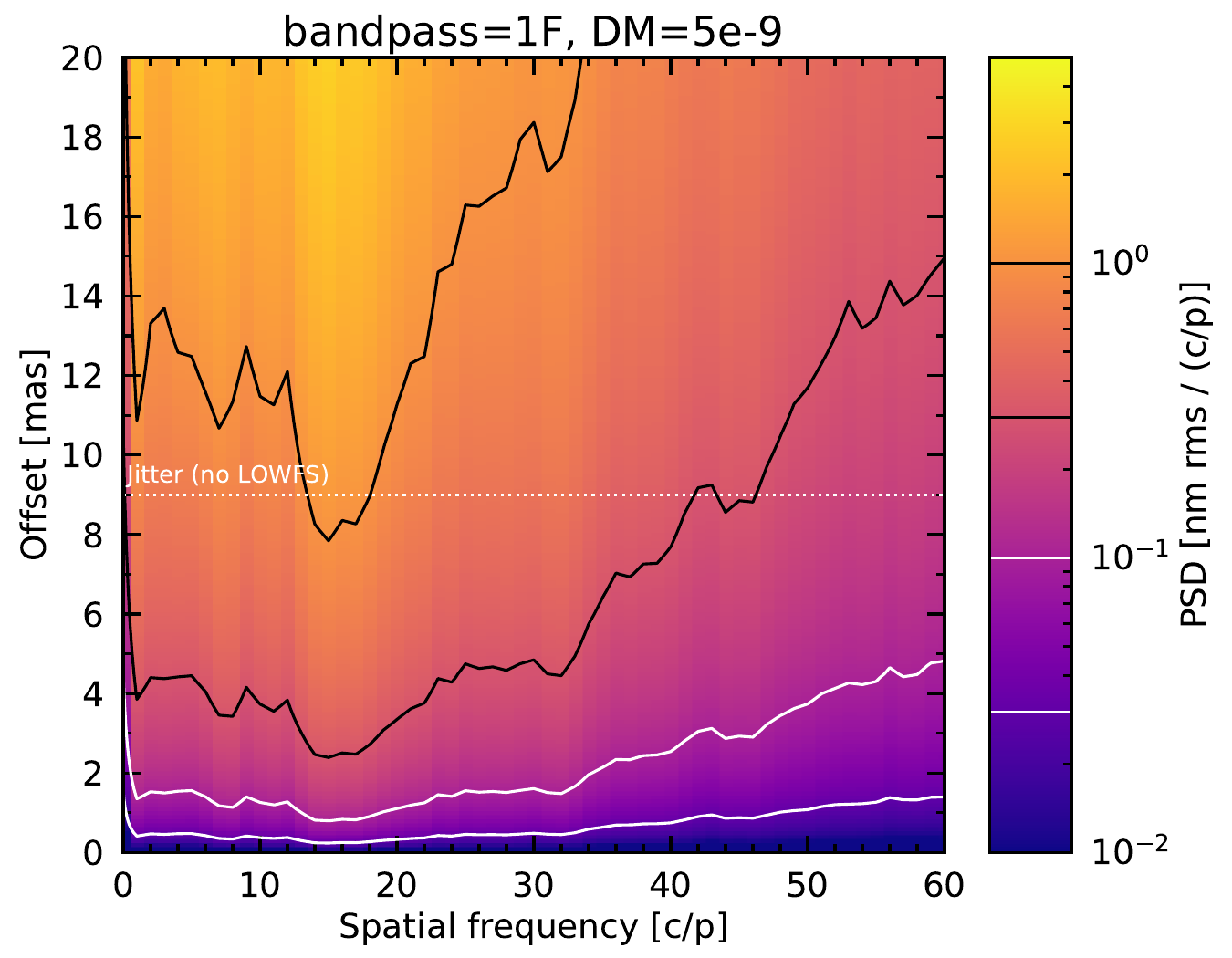}
  \caption{Same as Fig.~\ref{fig:offset_dm=flat} assuming a \dmconf DMs configuration.}
  \label{fig:offset_dm=5e-9}
\end{figure}

The goal of this simulation is to evaluate the impact of a pointing offset of the PSF on the ZWFS mask. If the PSF is not well centred, it impacts the dynamic of the sensor and the reconstruction, especially if a large part of the dynamic range is dominated by the tip and tilt induced by the offset on the mask.

The steps of the simulation are detailed below. The simulations are done in two DMs configuration, flat and \dmconf, and using the full band 1 filter (1F). All the data are generated without detector noise, to identify the fundamental limitations induced by the offset.

\begin{enumerate}
    \item A clear pupil image is generated by removing the mask in the simulation.
    \item ZWFS measurements are generated on a linear grid of offsets, from 0 to 20\,mas in steps of 0.1\,mas. The offsets are induced along a single direction (+x) because the results are symmetric with respect to the centre of the mask; it is therefore not necessary to explore the full 2d space.
    \item The OPD maps are reconstructed.
    \item The tip contribution is estimated and removed in the reconstructed OPD maps. Since we are artificially inducing an offset in +x, the tip is consequently induced on the wavefront, but we are in fact interested in higher order aberrations. Not removing the tip contribution would prevent a fair comparison between the ``reference'' OPD map obtained with a zero offset and the other maps obtained in presence of an offset.
    \item The final reconstruction error is obtained by subtracting the reference OPD map to the other OPD maps.
\end{enumerate}

Results are presented in Fig.~\ref{fig:offset_dm=flat} and \ref{fig:offset_dm=5e-9} for the flat and \dmconf DMs configurations, respectively. The figures show the reconstruction error over the full wavefront (in nm rms) as a function of the offset, as well as the decomposition of this reconstruction error in terms of spatial frequency (in nm rms / (\cpup)). The decomposition in spatial frequencies corresponds to the integration of the 2d power spectral density in fixed bins of spatial as defined by $\sigma(f)$ in Vigan et al. (2019)\cite{Vigan2019}. This ``integrated PSD'' is relatively easy to interpret as it corresponds to a certain level of wavefront aberrations at a given spatial frequency.

In the flat DMs configuration, assuming that the offset is small ($<$10\,mas), the reconstruction error is of the error of 1 to 2\,nm\,rms. It is interesting to note that it is mostly flat in terms of spatial frequency, which means that all spatial frequencies are affected in the same way by an offset. In the \dmconf DMs configuration, the situation is slightly worse, with a stronger impact at smaller offsets. In this configuration, the $\sim$1\,nm\,rms limit is reached for an offset of the order of 2\,mas. The impact is also less uniform in terms of spatial frequencies, with a smaller error on higher spatial frequencies.

This presentation of the reconstruction error with respect to an ideal case is probably slightly pessimistic since an analysis of the temporal evolution of the aberrations would be based on differential measurements done at different times. In this situation, a (small) static offset in tip and tilt is not necessarily a limitation except if the dynamic of the sensor is significantly decreased to a point where the differential aberration cannot be sensed any more. Nonetheless, it is interesting to see that a static offset of a few mas would prevent an absolute reconstruction at better than the nanometre.

\subsection{Impact of jitter}
\label{sec:perf:jitter}

\begin{figure}
  \centering
  \includegraphics[width=0.49\textwidth]{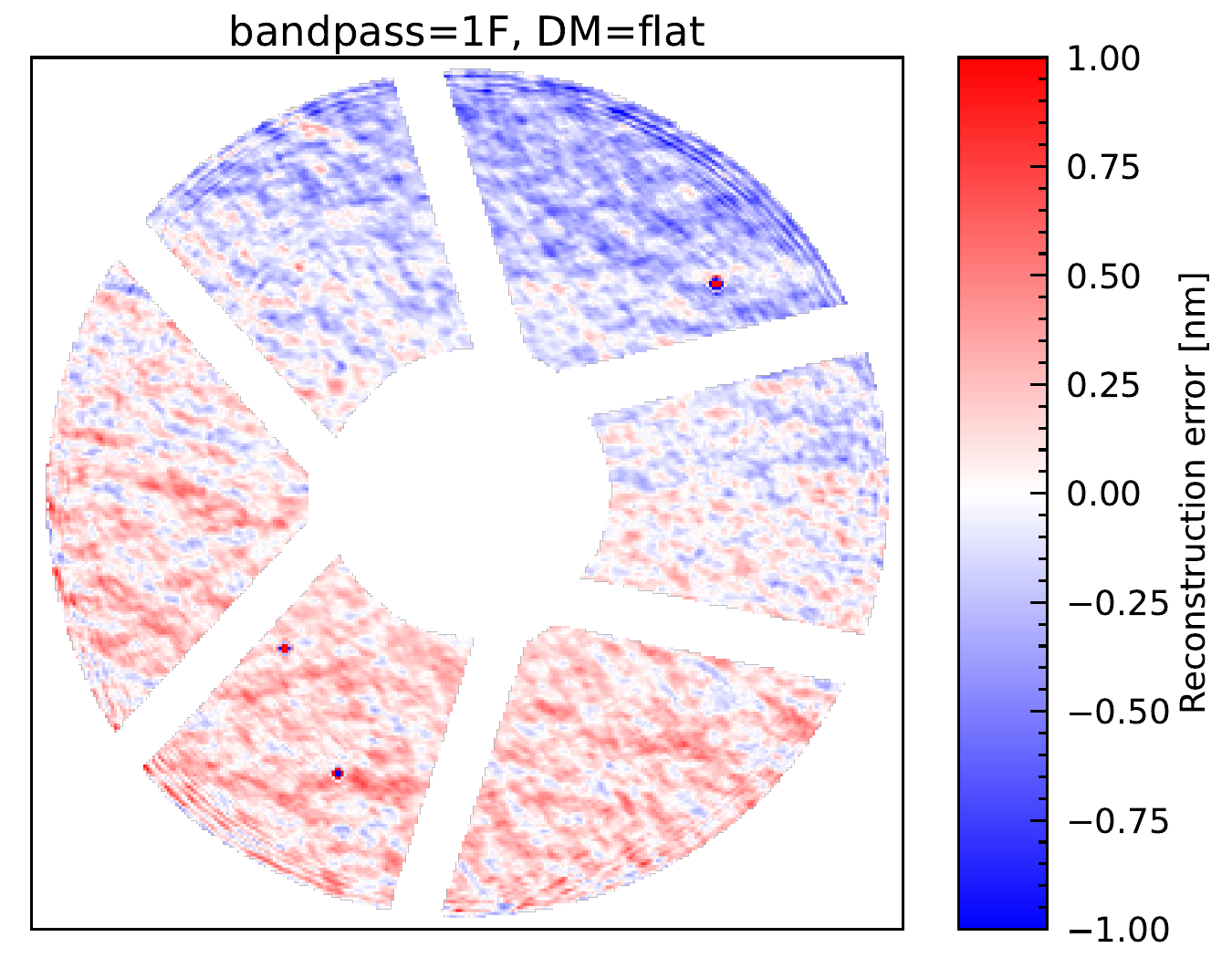}
  \includegraphics[width=0.49\textwidth]{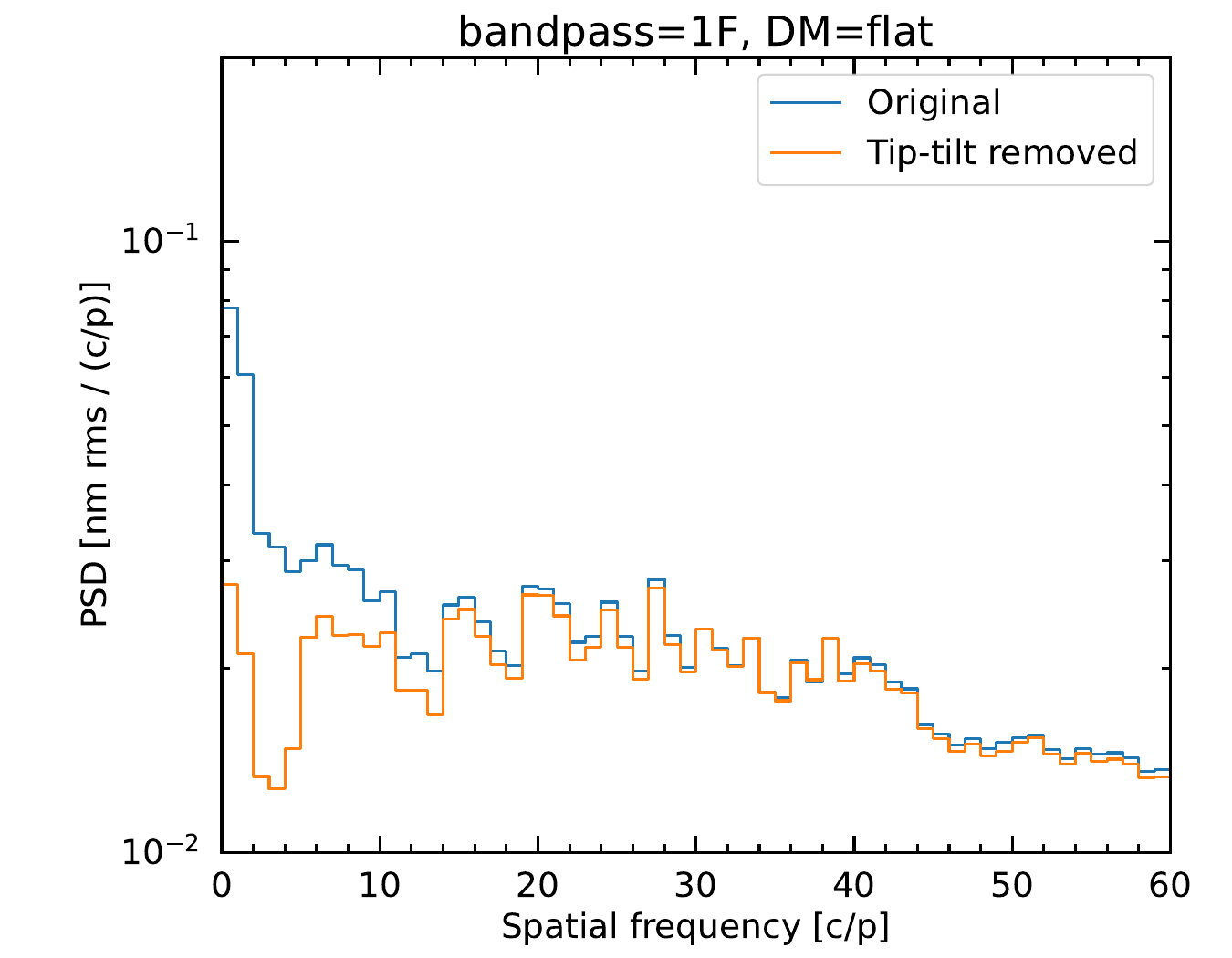}
  \caption{Wavefront reconstruction error with the ZWFS assuming an observatory jitter of 9\,mas and no LOWFS correction, assuming a flat DMs configuration. The left plot shows the full wavefront reconstruction error (in nm rm), while the right plot shows the decomposition in terms of spatial frequency (in nm rms / (\cpup)) without and with tip-tilt removal.}
  \label{fig:jitter_dm=flat}
\end{figure}

\begin{figure}
  \centering
  \includegraphics[width=0.49\textwidth]{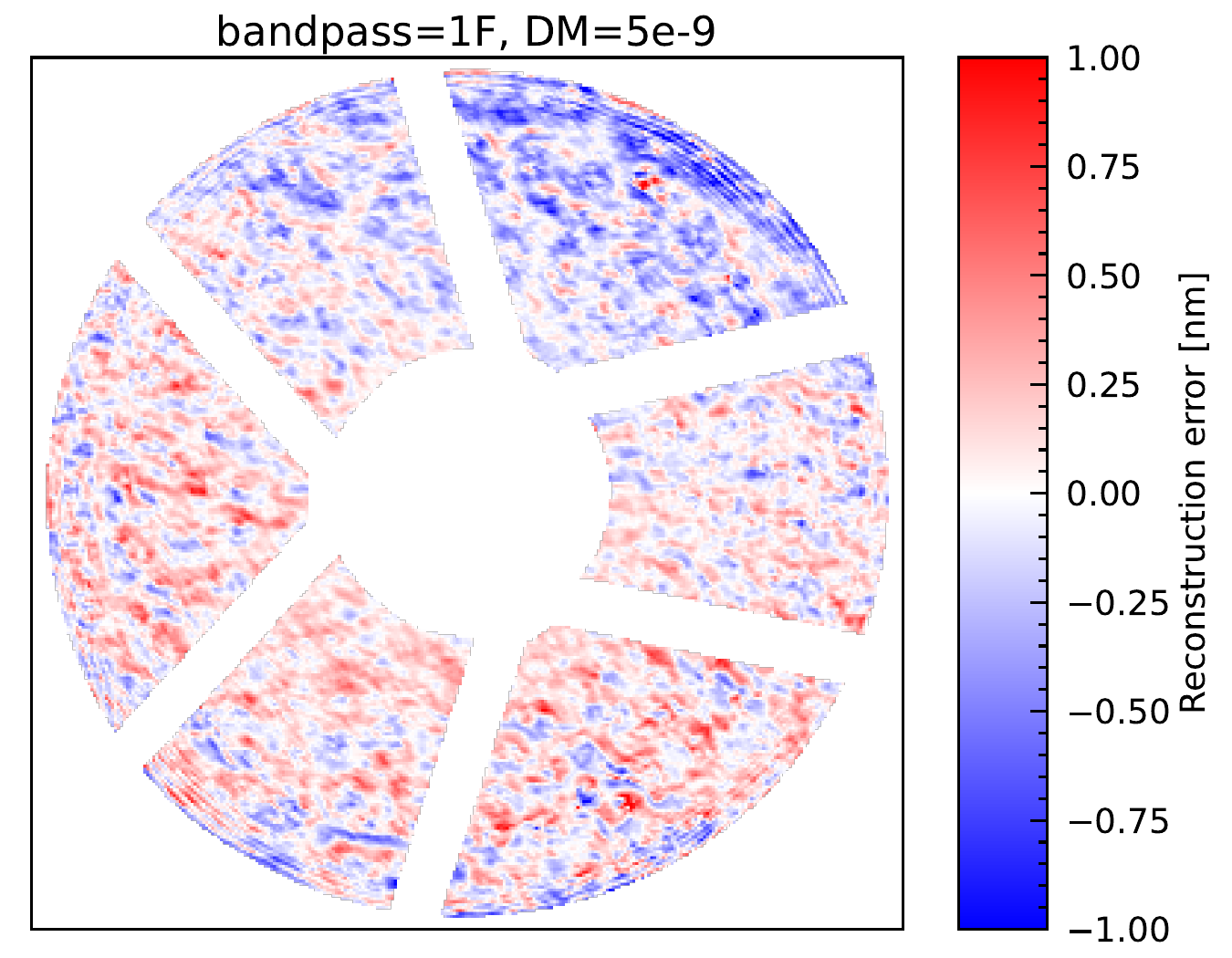}
  \includegraphics[width=0.49\textwidth]{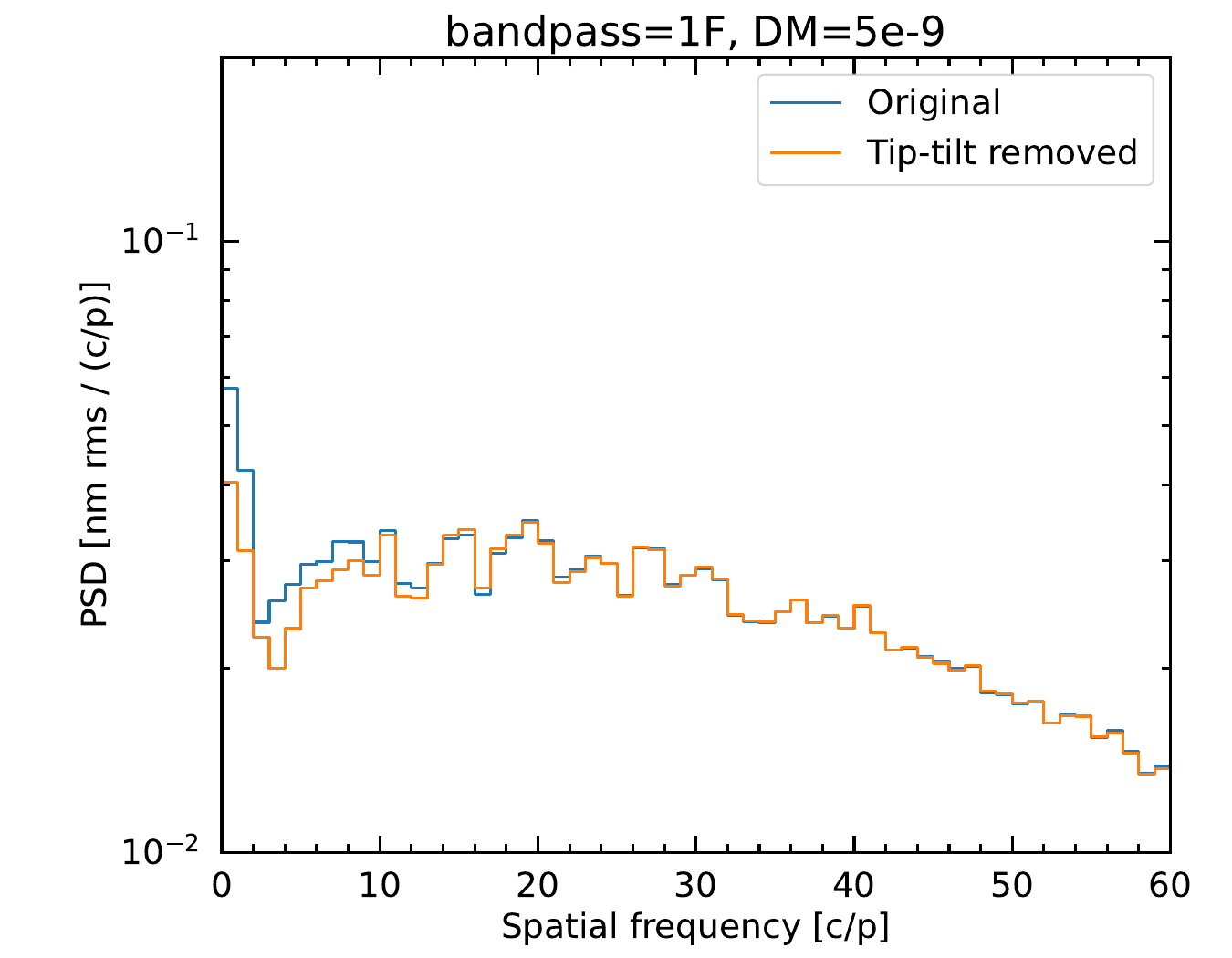}
  \caption{Same as Fig.~\ref{fig:jitter_dm=flat} assuming a \dmconf DMs configuration.}
  \label{fig:jitter_dm=5e-9}
\end{figure}

A second important limit is the jitter, specifically in the case where the LOWFS could not work in closed-loop with the dual-path ZWFS. There are good hints from internal JPL studies that the LOWFS should be able to work in this configuration, but since this has not been demonstrated in TVAC there are still some unknowns. With LOWFS, the jitter of the PSF is of the order of 0.4\,mas\,rms on $V<5$ stars, while without LOWFS the observatory jitter will dominate and would be of the order of the order of 9\,mas\,rms. The final observatory jitter once in flight might actually be better, but this will not be confirmed before launch and start of operations. In the present simulation we therefore assume a jitter of 9 mas during the ZWFS measurements.

In the simulation we generate ZWFS measurements with and without jitter, and we compare the reconstructed OPDs. Both simulations assume that the PSF is perfectly centred on the mask (i.e., jitter is a random perturbation around the centre of the mask). Again the simulations are done in two DMs configuration, flat and \dmconf, and using the full band 1 filter (1F). All the data are generated without detector noise.

Results are presented in Fig.~\ref{fig:jitter_dm=flat} and \ref{fig:jitter_dm=5e-9} for the flat and \dmconf DMs configurations, respectively. In both configurations, the impact of the jitter is small, with an estimated error of less than 3\,nm\,rms/(\cpup). The dominant term is a mix of tip and tilt, which is unexpected since the jitter is simulated as a distribution of PSF positions around the center of the mask. When removing the tip-tilt contribution, the remaining distribution is mostly flat in the flat DMs configuration, and has a downward trend at higher spatial frequencies for the \dmconf configuration. Overall, we conclude that the impact of the jitter is not negligible but still at a low level. If the LOWFS loop cannot be closed with the dual-path ZWFS, we would still be able to perform meaningful differential OPD measurements.

\subsection{Sensitivity to small phase errors}
\label{sec:perf:small_errors}

\begin{figure}[!h]
  \centering
  \includegraphics[width=0.49\textwidth]{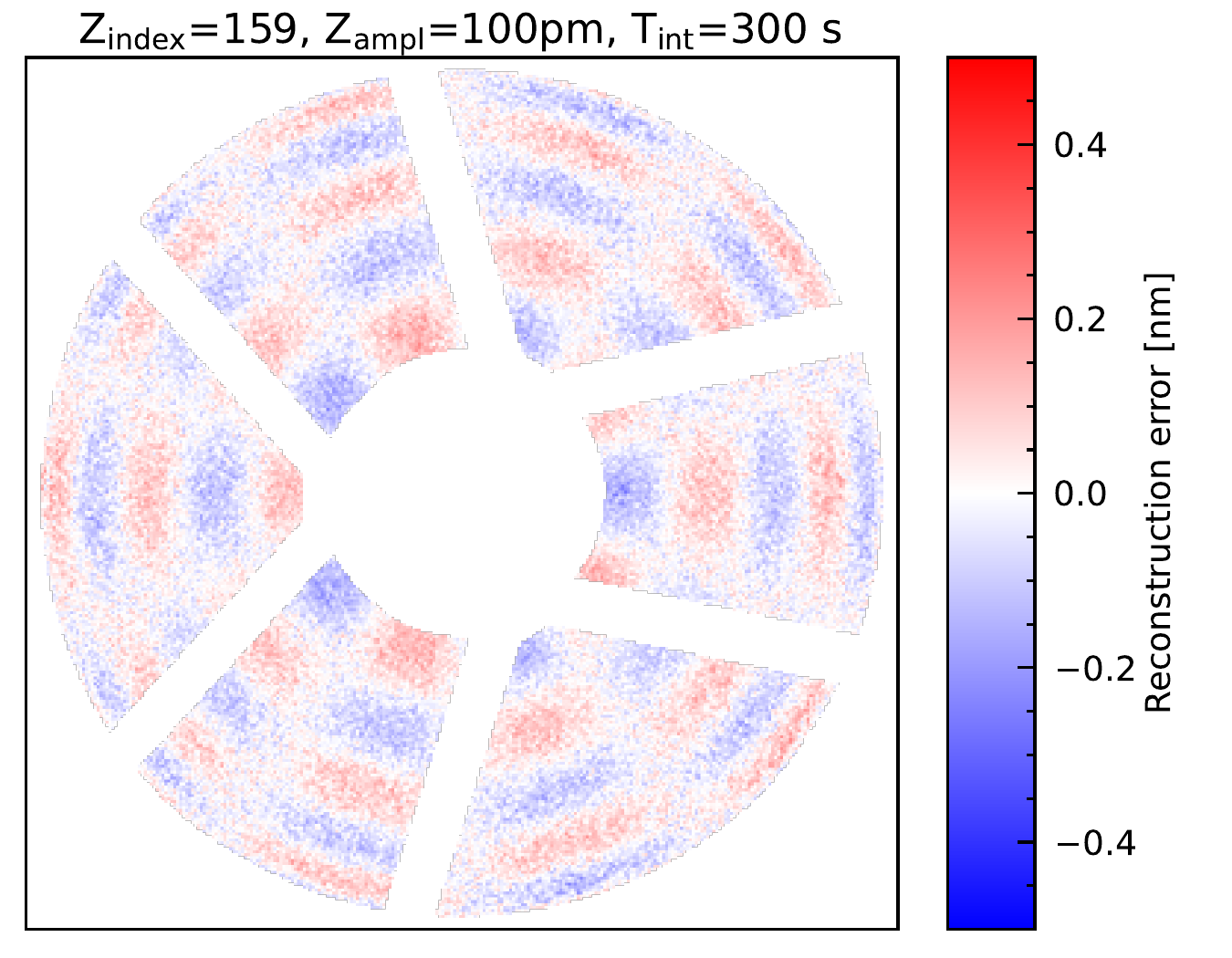}
  \caption{Reconstructed differential OPD map based on two exposures of 300\,s containing a high-order differential aberration of 100\,pm in Z159. The exposures are simulated on a $V = 2$ star using an EMCCD gain of 1 and individual integration times of 2\,s per exposure. The integrated PSD of the reconstruction is provided in the left plot of Fig.~\ref{fig:exposures_psd_Tint=all} (green line).}
  \label{fig:exposure_z159_100pm_300s}
\end{figure}

\begin{figure}
  \centering
  \includegraphics[width=0.49\textwidth]{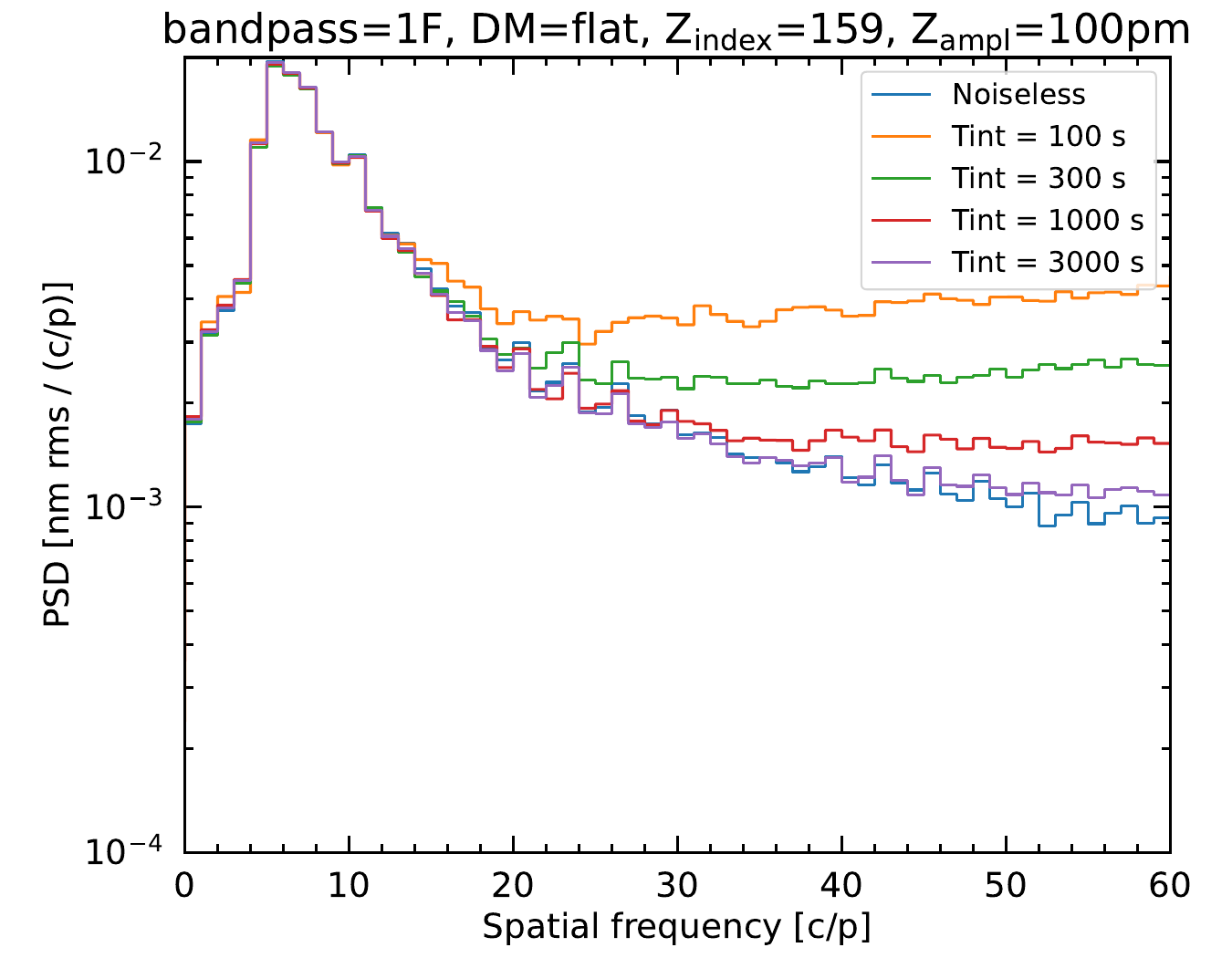}
  \includegraphics[width=0.49\textwidth]{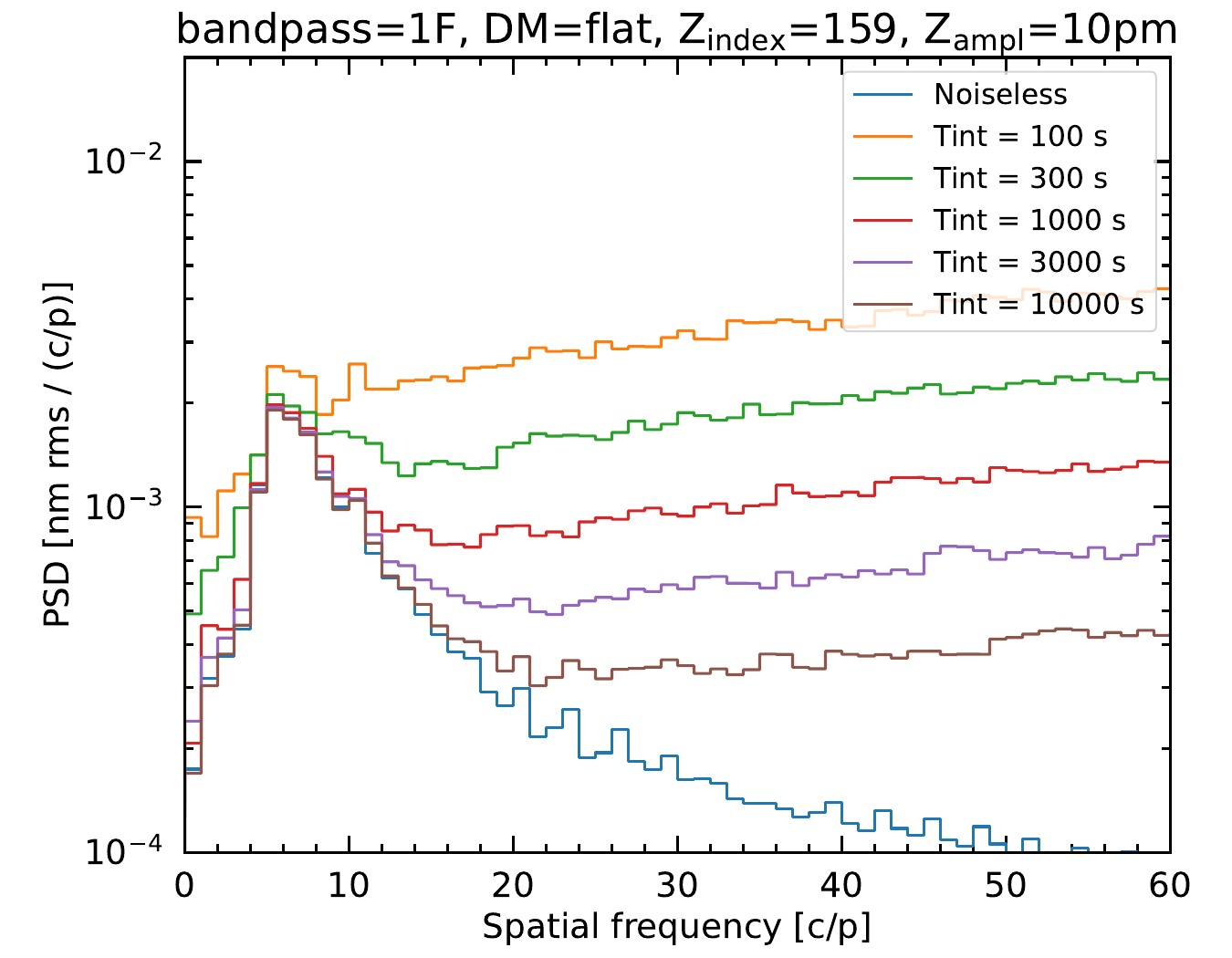}
  \caption{Integrated PSD of the reconstructed differential aberration assuming two levels of differential aberration (100\,pm and 10\,pm on the left and right, respectively) and different total integration times from 100\,s to 10\,000\,s.}
  \label{fig:exposures_psd_Tint=all}
\end{figure}

Previous simulations have all been done without including the EMCCD effects and associated noises. In this section we explore the sensitivity of the dual-path ZWFS from realistic exposures. Since ZWFS measurements are intended for calibrations and characterisation of the observatory and Coronagraph, we simulate exposures assuming a bright star ($V = 2$) to maximise the signal-to-noise ratio (SNR). With such a bright target, exposures of 2\,s with an EMCCD gain of 1 allow reaching $\sim$70\% of the full-well capacity (but with no saturation) on ZWFS pupil images.

We perform simulations where we introduce a differential aberration between two series of measurements and simulate different total exposure times. We intentionally select a high-order Zernike polynomial, $Z = 159$ (Noll numbering), that is much higher than what the LOWFS can sense. We simulate data without aberration and with a differential aberration, we reconstruct independently the OPD maps, and finally we subtract the two OPD maps to measure only the differential aberration. To assess then sensitivity of the dual-path ZWFS, we perform comparisons between a noiseless simulation and simulations with simulated observations at different total integration times. Figure~\ref{fig:exposure_z159_100pm_300s} shows an example of such a differential measurement assuming an initial 100\,pm differential aberration and a total integration of 300\,s for each of the two exposures.

Figure~\ref{fig:exposures_psd_Tint=all} explores more exhaustively the quality of the reconstruction assuming two levels of differential aberrations, 100\,pm and 10\,pm, and increasing total integration time for the individual exposures, with and without the differential aberration. A first finding is that a differential aberration of the order of 100\,pm can be accurately sensed in a few minutes of total integration time up to 20 or 30\,\cpup. If there is a specific requirement to sense the wavefront up to the spatial frequency set by the DM actuators, then an integration time of the order of the hour will be required. Of course, the longer the integration time the better the reconstruction: as expected, the noise floor in the measurements defines the lowest level of aberration that can be measured, and this limits decreases in $\frac{1}{\sqrt{N_{\mathrm{phot}}}}$.

Then, sensing aberrations of the order of 10\,pm is a totally different regime with the current configuration of the dual-path ZWFS. At the level of 10\,pm differential aberration, only the lowest spatial frequencies ($<$10\,\cpup) can be accurately reconstructed with integration times of the order of one hour. This relatively poor performance for small aberrations is, in part, related to the oversized pupil image on the EMCCD science detector: the pupil image is significantly over-sampled, with almost 300 pixels over the diameter of the pupil. While this configuration can be interesting to be understood the instrument, it spreads the useful signal over more pixels than strictly required to sense and maintain the spatial frequencies of interest for the DH.

\section{OS11 time series}
\label{sec:os11}

Observing Scenario (OS) 11 is a simulated long sequence of observations that has been designed to explore the limits of the Roman Coronagraph. All the details are provided \href{https://roman.ipac.caltech.edu/page/coronagraph-public-images-html}{here}. The sequence is based on a reference star with $V = 2$ and a science target with $V = 5$. The sequence includes a DH digging on the reference star, followed by a slew to the science target, and then a long series of observations on the science target. The sequence is designed to explore the evolution of the DH over time, and to assess the need for DH touch-ups during the long science observations. The OS couples an optical propagation through the Coronagraph with a realistic thermal and mechanical model of the full observatory. The OS11 sequence is therefore a good test case to explore the performance of the dual-path ZWFS in a realistic scenario.

\begin{figure}
  \centering
  \includegraphics[width=0.6\textwidth]{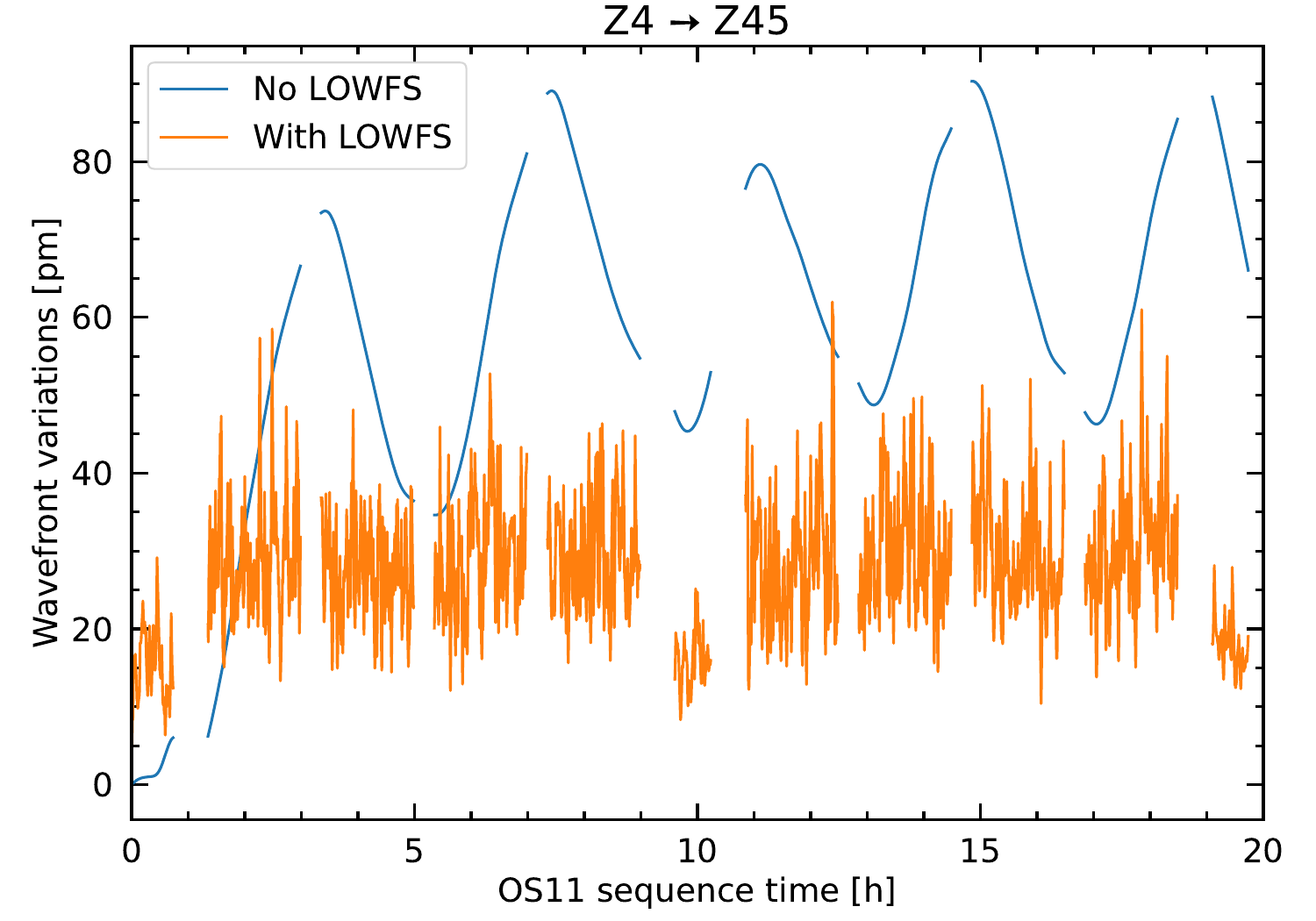}
  \caption{Low-order aberrations (Z4 to Z45) in the OS11 sequence, without (blue) and with (orange) the LOWFS correction. The LOWFS correction is applied only on Z4 to Z11. The holes in the sequence correspond to the slew from the reference star to the science target, to the observatory rolls and to the DH touch-ups. The LOWFS correction is applied every 10\,s with a sensing bandwidth of 100\,s.}
  \label{fig:os11_sequence}
\end{figure}

Among the distributed OS11 data products are the input low-order aberrations projected over Z4 to Z45. There cumulated effect on the wavefront is presented in Fig.~\ref{fig:os11_sequence}. The figure shows the evolution of the wavefront error (WFE) in nm rms over the full OS11 sequence. The WFE is dominated by low-order aberrations, which are well within the capabilities of the LOWFS to correct. However, there are also higher order aberrations that evolve over time, which could be sensed by the dual-path ZWFS. The figure also shows the remaining WFE after the LOWFS correction over Z4 to Z11. The LOWFS model in the OS11 sequence uses the same algorithms as flight for sensing and deriving corrections. The LOWFS-measured are Zernikes reported every 10 s with 100\,s sensing bandwidth. The defocus (Z4) is controlled with the Focus Correction Mechanism, while Z5-Z11 are controled with DM1.

\begin{figure}
  \centering
  \includegraphics[width=1\textwidth]{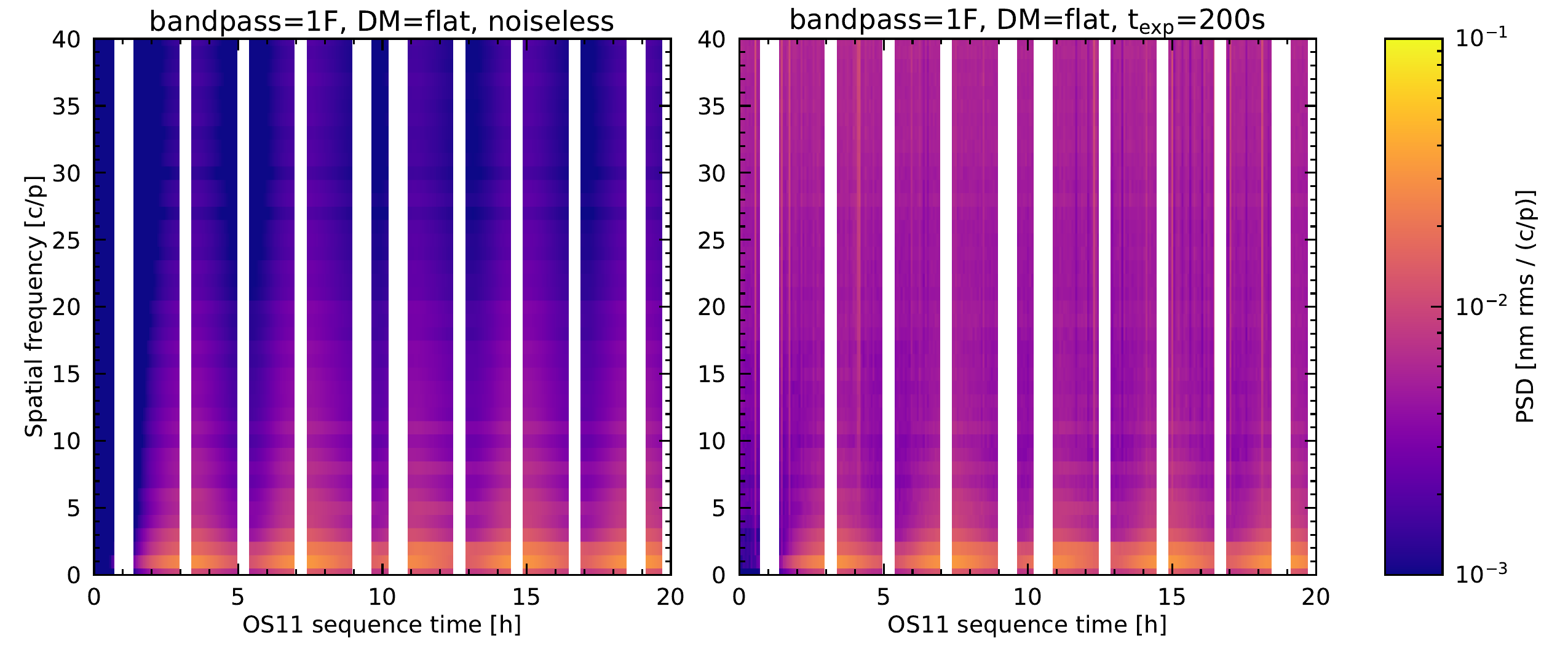}
  \caption{Integrated PSD of the OPD maps reconstructed for the OS11 sequence with no LOWFS and a sampling every 250\,s. The left plot corresponds to the reconstruction with no noise, while the right plot corresponds to the reconstruction assuming integration times of 200\,s for the dual-path ZWFS measurements.}
  \label{fig:os11_sampling=250_lowfs=0_texp=200_2d}
\end{figure}

We use this input sequence os Zernike coefficients as input to generate dual-path ZWFS measurements in the same way as described previously. The sequence is sampled either every 250 or every 1000\,s, with total integration times of 200\,s and 800\,s for the ZWFS exposures, respectively. We then compute the integrated PSDs of the OPD maps and compare them to an ideal reconstruction without noise. In Fig.~\ref{fig:os11_sampling=250_lowfs=0_texp=200_2d} we present an example for the sequence using 200\,s of integration time for the ZWFS and assuming the LOWFS loop is open. In these initial simulations we make the hypothesis that the WFE remains static over the course of the exposures. While this is obviously not correct, the idea of these simulations is to estimate the potential of the dual-path ZWFS to sense the evolution of the WFE over time, and to identify the limitations of the sensor in this context.

\begin{figure}
  \centering
  \includegraphics[width=1.0\textwidth]{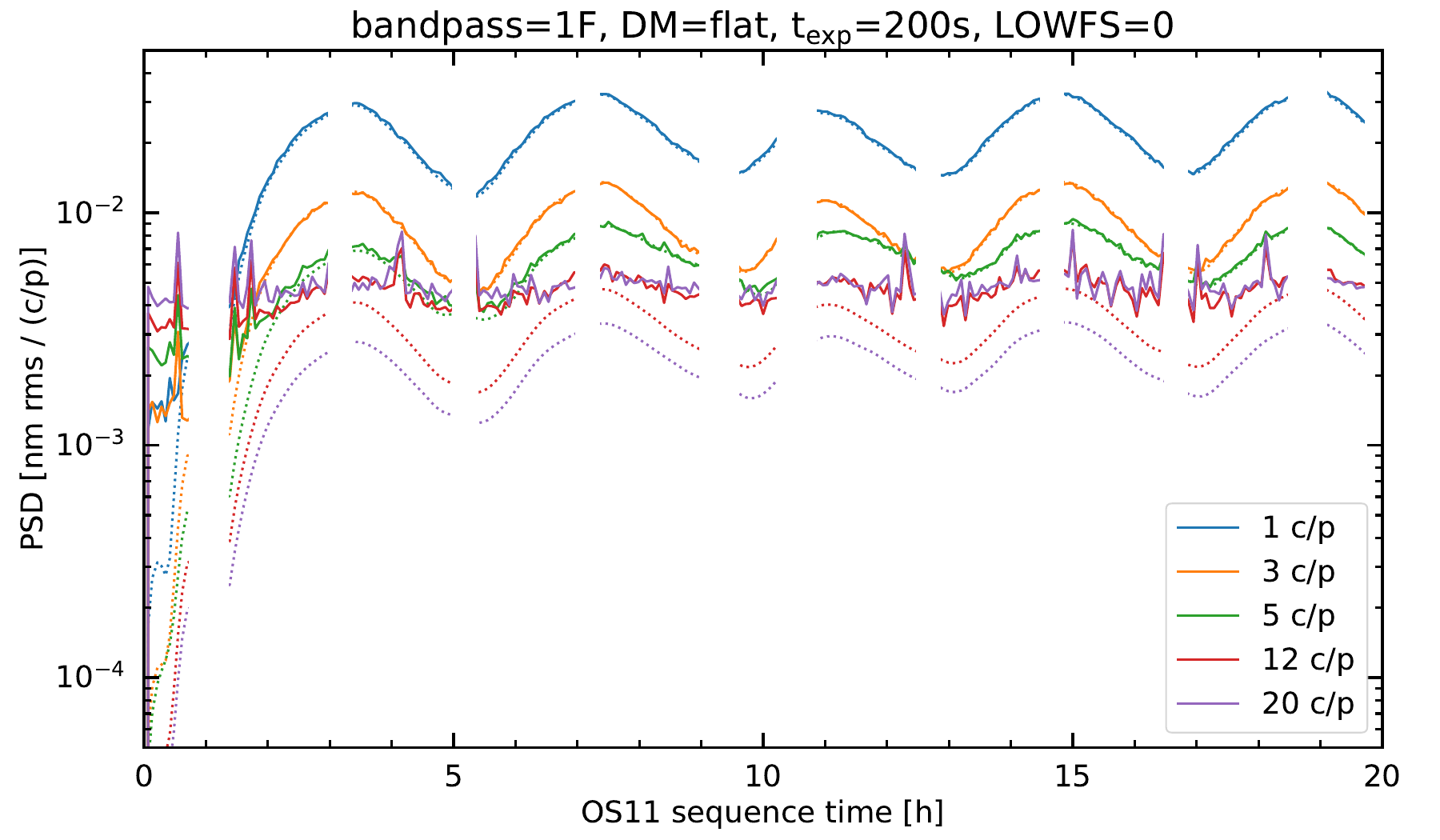}
  \includegraphics[width=1.0\textwidth]{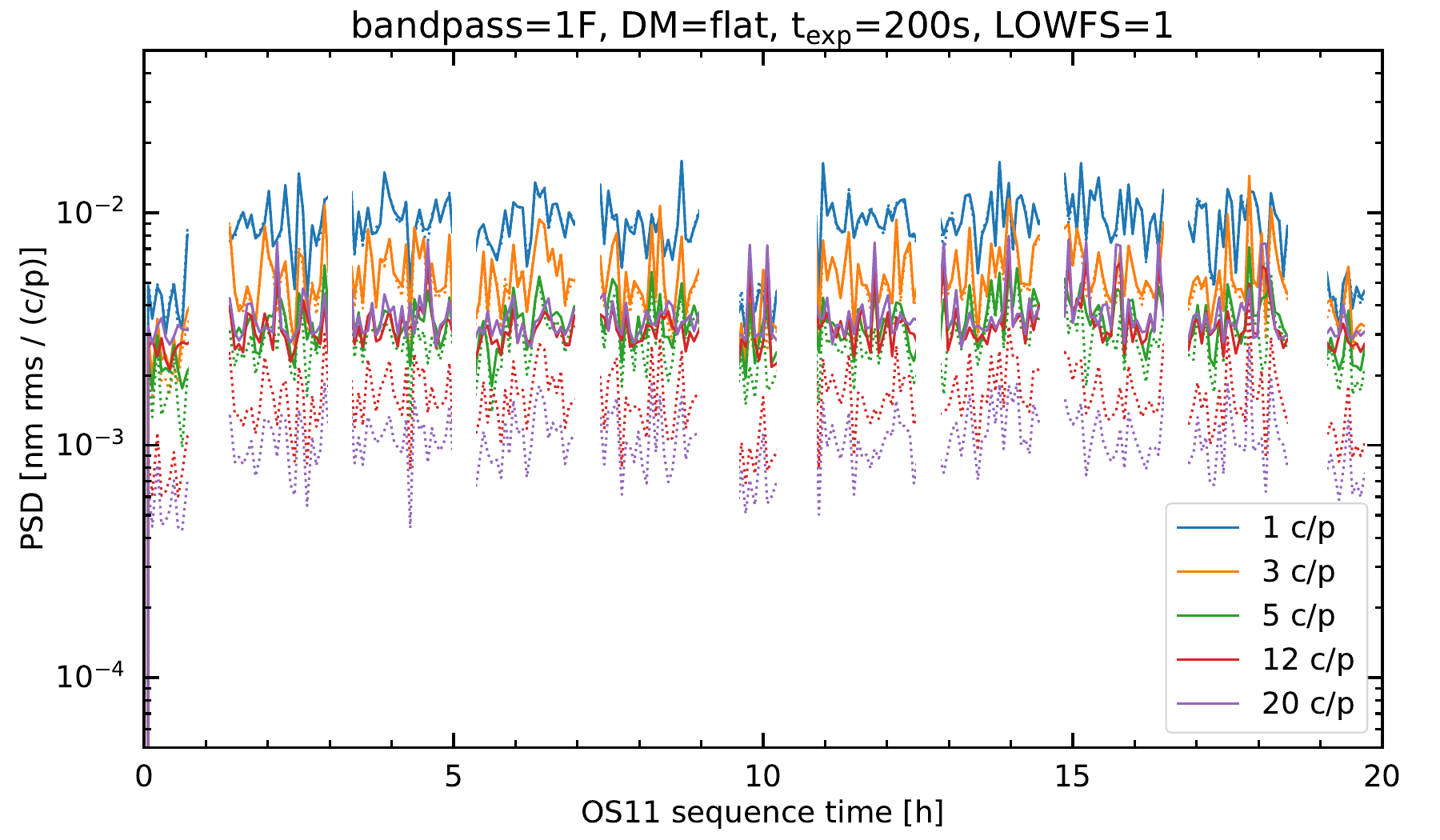}
  \caption{Wavefront error reconstruction at five different spatial frequencies based on dual-path ZWFS measurements simulated for an OS11-like sequence. The plot compares the ideal noiseless reconstruction (dotted line) to a reconstruction based on measurements with 200\,s ZWFS measurements. The top plot is a simulation in a configuration without the LOWFS, while the bottom plot includes the LOWFS correction of the low orders (Z4 to Z11).}
  \label{fig:os11_sampling=250_texp=200}
\end{figure}

\begin{figure}
  \centering
  \includegraphics[width=1.0\textwidth]{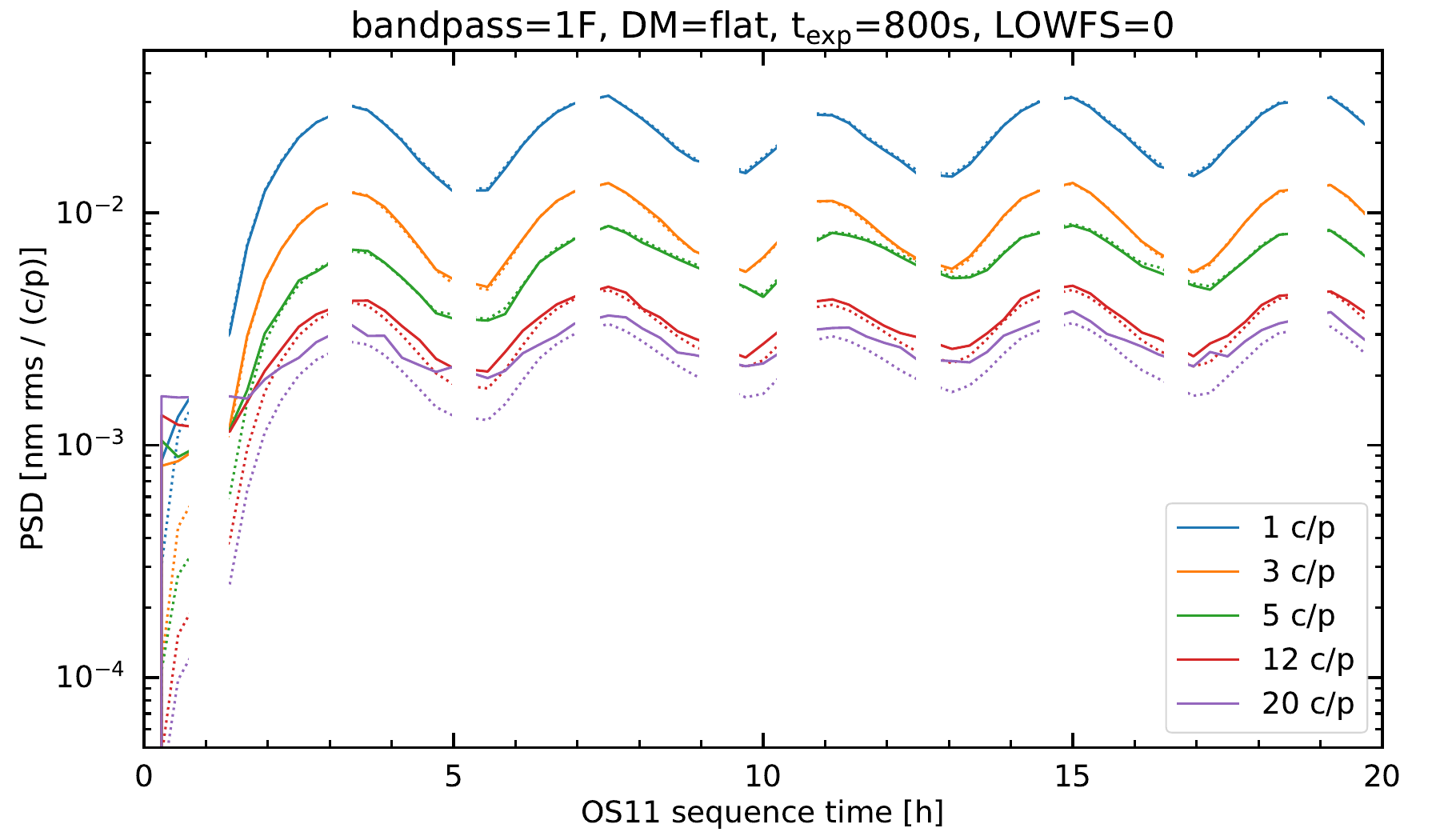}
  \includegraphics[width=1.0\textwidth]{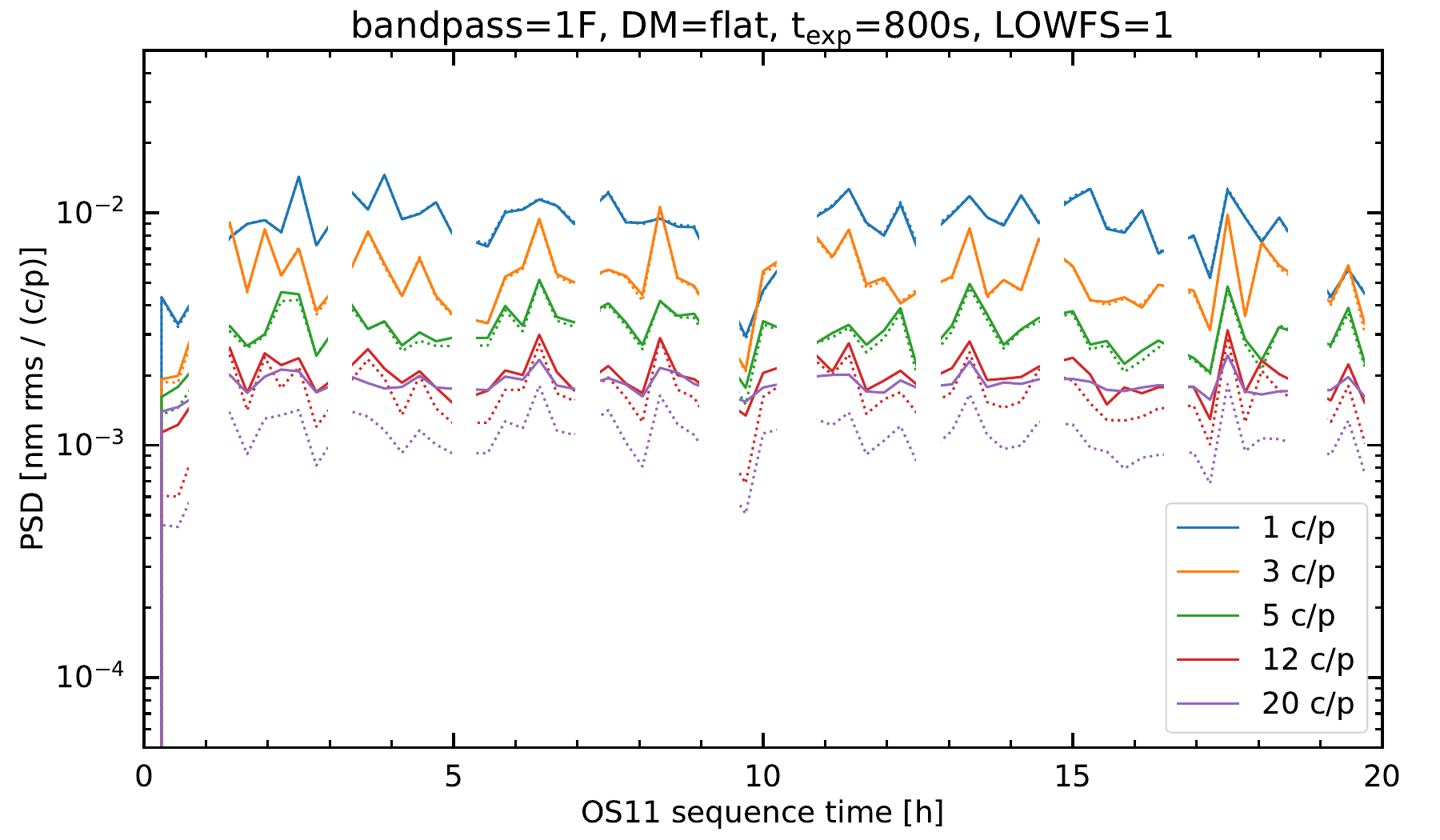}
  \caption{Same as Fig.~\ref{fig:os11_sampling=250_texp=200} but for a sampling every 1000\,s and exposure times of 800\,s for the dual-path ZWFS measurements.}
  \label{fig:os11_sampling=1000_texp=800}
\end{figure}

The results are presented in Fig.~\ref{fig:os11_sampling=250_texp=200} and \ref{fig:os11_sampling=1000_texp=800} for 200\,s and 800\,s integration times, respectively. These plots show the reconstruction comparison between the noiseless and noisy cases at five different spatial frequencies (1, 3, 5, 12 and 20\,\cpup). They also present results without LOWFS and with the LOWFS correction.

Whatever the integration time, the effect of the LOWFS is clearly visible, with a significant attenuation of the low spatial frequencies. In the configuration without LOWFS, there is no compensation of the variation of the low spatial frequency aberrations, and the wavefront error variations are much higher throughout the OS11 sequence. At the low integration time (200\,s), the spatial frequencies up to 5\,\cpup are well reconstructed, but there is a change of regime between 5 and 12\,\cpup. At 12\,\cpup and beyond, the reconstruction is no longer accurate anymore due to the noise in individual ZWFS measurements. Increasing the exposure time of the ZWFS measurements to 800\,s improves the situation, providing an accurate reconstruction up to $\sim$10\,\cpup. At 12\,\cpup, and even 20\,\cpup, the reconstruction still contains meaningful information but there reconstruction error increases. In the configuration with the LOWFS compensation, most low spatial frequency aberrations are reduced below $10^{-2}$\,nm\,rms / (\cpup). Here again, with the short exposure time we are able to obtain a good reconstruction up to $\sim$5\,\cpup, but beyond that the measurements are noisy. Similarly, with the longer integration time the reconstruction is relatiely accurate up to 10\,\cpup.

\section{Conclusions and perspectives}
\label{sec:conclusions}

The ZWFS is now a well established technique for high-contrast imaging applications in space. The Roman coronagraph will be the very first instrument to put it in practice within its low-order wavefront sensor to maintain a low-level of aberrations during long coronagraphic observations. However, the filtered nature of the LOWFS ZWFS does not allow sensing aberrations at spatial frequencies higher than a few cycles/pupil. Although they are expected to be at very low levels (few pm), it could still be interesting to be able to measure them and understand their variations with time\cite{Vigan2022}. For this application, the dual-path ZWFS included in the FPAM is most suited as it provide a very high spatial sampling, with almost 300 pixels over the diameter of the pupil.

In the present work we have investigated the performance of the dual-path ZWFS using the \texttt{corgisim} tool developed by the CPP. The simulations show that the sensor is, in theory, sensitive to picometer-level aberrations in an ideal noiseless case. One of the still open questions for the Roman dual-path ZWFS is whether the LOWFS will be able to work in closed-loop with the sensor in place. Although unpublished work at JPL shows that it should, the sensor could not be tested in TVAC, so there is no definitive answer. With that in mind, we performed noiseless simulations that show that the impact of a small offset of the PSF on the phase dimple due to a misalignment in the acquisition and centering remains manageable if the offset is of the order of a couple of milli-arcseconds. Moreover, the effect of the jitter of the PSF assuming that the LOWFS loop is not active also has a limited impact on the phase reconstruction, with mostly a tip-tilt residual term.

One of potential use of the dual-path ZWFS is the acquisition of time series measurements to monitor the variation of aberrations introduced by the observatory and the instrument up to mid- and high-spatial frequencies. Such an analysis would be based on differential measurements obtained at different times. We therefore investigated the sensitivity of the dual-path ZWFS with realistic, noisy simulations generated with \texttt{corgisim} and applying the EMCCD detection noises. Assuming observation of a bright star ($V = 2$) and the band 1 broadband filter, we introduced differential aberrations of 100\,pm and 10\,pm on $Z_{159}$, and simulated observations from 100\,s up to 10\,000\,s. At 100\,pm-level, the dual-path ZWFS should be able to properly sense aberrations up to 20\,\cpup in 5\,min and up to 50-60\,\cpup in slightly less than an hour. The situation is obviously much worse for 10\,pm-level aberrations, where close to 3\,h of integration time would be necessary to sense accurately aberrations up of $\sim$15\,\cpup.

Simulations based on the OS11 observing scenario with and without the LOWFS demonstrate the ability to sense low-order aberrations, or their residuals if the LOWFS loop is active, at a cadence of $\sim$200\,s. However, this only concerns the lowest spatial frequencies that can already be measured by the LOWFS. For higher spatial frequencies, beyond 10 or 15\,\cpup, integration times much longer than the typical timescale of wavefront variations would probably be required, which decreases the interest of a dedicated monitoring with the dual-path ZWFS.

Still, obtaining one long monitoring sequence with the dual-path ZWFS would still be of high interest for the future. First, the filtered nature of the LOWFS produces an incomplete and biased signal for spatial frequencies beyond a few cyles-per-pupil. Obtaining a sequence with joint measurements with the LOWFS and the dual-path ZWFS would be extremely valuable to understand the performance of the LOWFS and its associated control loop. Moreover, the dual-path ZWFS provides information up to high spatial frequencies. Even if variations beyond Z11 are expected to be at extremelly low levels, hardly measurable at short exposure times with the dual-path ZWFS, we cannot yet exclude some surprises once the observatory has been launched and commissioned. ZWFS have already demonstrated their value for monitoring and diagnosing high-contrast imaging instruments in the past, and the dual-path ZWFS would be a valuable solution for the Roman Coronagraph. On the longer term, HWO will almost certainly rely on ZWFS for its LOWFS. It is therefore important to gain as much knowledge with this sensor on Roman.

In practice, some work would certainly be needed to support the dual-path ZWFS in Roman. The first aspect would be to develop a simple and robust centering procedure. One of the difficulties is that the phase dimples of the ZWFS are completely transparent, which makes the centering procedure tricky. The seven dual-path ZWFS phase dimples are located in the HLC12 substrate of the FPAM (column 7, rows 2 to 8) exactly halfway between metal spots to the left and right\cite{Riggs2025}. There are also metal spots above and below in rows 1 and 9. The star could be aligned to the metal spots on the left and right of a ZWFS spot, and then, the FPAM stage could be driven to the middle position. Depending on the accuracy of this coarse centering, a fine centering procedure could then be performed, but we have seen in Sect.~\ref{sec:perf:offset} that a small offset on the mask is not necessarily a major limitation, especialy if the LOWFS loop can work, which will also need to be demonstrated on-sky.

\section{Acknowledgements}

Financial support for this study was provided by the Centre National d'Études Spatiales (CNES), France (ROR: \url{https://ror.org/04h1h0y33}) within the framework of the Roman Space Telescope space mission. Part of this work was carried out at the Jet Propulsion Laboratory, California Institute of Technology, under a contract with the National Aeronautics and Space Administration (80NM0018D0004).

\bibliography{biblio} 

@ARTICLE{Vigan2019,
       author = {{Vigan}, A. and {N'Diaye}, M. and {Dohlen}, K. and {Sauvage}, J.-F. and {Milli}, J. and {Zins}, G. and {Petit}, C. and {Wahhaj}, Z. and {Cantalloube}, F. and {Caillat}, A. and {Costille}, A. and {Le Merrer}, J. and {Carlotti}, A. and {Beuzit}, J.-L. and {Mouillet}, D.},
        title = "{Calibration of quasi-static aberrations in exoplanet direct-imaging instruments with a Zernike phase-mask sensor. III. On-sky validation in VLT/SPHERE}",
      journal = {\aap},
         year = 2019,
        month = sep,
       volume = {629},
          eid = {A11},
        pages = {A11},
          doi = {10.1051/0004-6361/201935889},
archivePrefix = {arXiv},
       eprint = {1907.11241},
 primaryClass = {astro-ph.IM},
       adsurl = {https://ui.adsabs.harvard.edu/abs/2019A&A...629A..11V}
}

@ARTICLE{Vigan2022,
       author = {{Vigan}, A. and {Dohlen}, K. and {N'Diaye}, M. and {Cantalloube}, F. and {Girard}, J.~H. and {Milli}, J. and {Sauvage}, J.-F. and {Wahhaj}, Z. and {Zins}, G. and {Beuzit}, J.-L. and {Caillat}, A. and {Costille}, A. and {Le Merrer}, J. and {Mouillet}, D. and {Tourenq}, S.},
        title = "{Calibration of quasi-static aberrations in exoplanet direct-imaging instruments with a Zernike phase-mask sensor. IV. Temporal stability of non-common path aberrations in VLT/SPHERE}",
      journal = {\aap},
         year = 2022,
        month = apr,
       volume = {660},
          eid = {A140},
        pages = {A140},
          doi = {10.1051/0004-6361/202142635},
archivePrefix = {arXiv},
       eprint = {2202.10470},
 primaryClass = {astro-ph.IM},
       adsurl = {https://ui.adsabs.harvard.edu/abs/2022A&A...660A.140V}
}

@MISC{pyZELDA,
       author = {{Vigan}, A. and {N'Diaye}, M.},
        title = "{pyZELDA: Python code for Zernike wavefront sensors}",
 howpublished = {Astrophysics Source Code Library, record ascl:1806.003},
         year = 2018,
        month = jun,
          doi = {10.5281/zenodo.6564082},
          eid = {ascl:1806.003},
archivePrefix = {ascl},
       eprint = {1806.003},
       adsurl = {https://ui.adsabs.harvard.edu/abs/2018ascl.soft06003V}
}

@ARTICLE{Zernike1934,
       author = {{Zernike}, F.},
        title = "{Diffraction theory of the knife-edge test and its improved form, the phase-contrast method}",
      journal = {\mnras},
         year = 1934,
        month = mar,
       volume = {94},
        pages = {377-384},
          doi = {10.1093/mnras/94.5.377},
       adsurl = {https://ui.adsabs.harvard.edu/abs/1934MNRAS..94..377Z}
}

@ARTICLE{Cady2025,
       author = {{Cady}, Eric and {Bowman}, Nicholas and {Greenbaum}, Alexandra Z. and {Ingalls}, James G. and {Kern}, Brian and {Krist}, John and {Marx}, David and {Poberezhskiy}, Ilya and {Eldorado Riggs}, A.~J. and {Ruane}, Garreth and {Seo}, Byoung-Joon and {Shi}, Fang and {Zhou}, Hanying},
        title = "{High-order wavefront sensing and control for the Roman Coronagraph Instrument (CGI): architecture and measured performance}",
      journal = {Journal of Astronomical Telescopes, Instruments, and Systems},
         year = 2025,
        month = apr,
       volume = {11},
          eid = {021408},
        pages = {021408},
          doi = {10.1117/1.JATIS.11.2.021408},
archivePrefix = {arXiv},
       eprint = {2507.23738},
 primaryClass = {astro-ph.IM},
       adsurl = {https://ui.adsabs.harvard.edu/abs/2025JATIS..11b1408C}
}

@ARTICLE{Krist2023,
       author = {{Krist}, John E. and {Steeves}, John B. and {Dube}, Brandon D. and {Eldorado Riggs}, A.~J. and {Kern}, Brian D. and {Marx}, David S. and {Cady}, Eric J. and {Zhou}, Hanying and {Poberezhskiy}, Ilya Y. and {Baker}, Caleb W. and {McGuire}, James P. and {Nemati}, Bijan and {Kuan}, Gary M. and {Mennesson}, Bertrand and {Trauger}, John T. and {Saini}, Navtej S. and {Rafels}, Sergi Hildebrandt},
        title = "{End-to-end numerical modeling of the Roman Space Telescope coronagraph}",
      journal = {Journal of Astronomical Telescopes, Instruments, and Systems},
         year = 2023,
        month = oct,
       volume = {9},
          eid = {045002},
        pages = {045002},
          doi = {10.1117/1.JATIS.9.4.045002},
archivePrefix = {arXiv},
       eprint = {2309.16012},
 primaryClass = {astro-ph.IM},
       adsurl = {https://ui.adsabs.harvard.edu/abs/2023JATIS...9d5002K}
}

@ARTICLE{Seo2025,
       author = {{Seo}, Byoung-Joon and {Fathpour}, Nanaz and {Arndt}, David and {Shields}, Joel and {Boussalis}, Dhemetrios and {Cady}, Eric and {Kern}, Brian},
        title = "{Demonstration of low-order wavefront sensing and control system of Roman coronagraph instrument in thermal vacuum testing environment}",
      journal = {Journal of Astronomical Telescopes, Instruments, and Systems},
         year = 2025,
        month = apr,
       volume = {11},
          eid = {021411},
        pages = {021411},
          doi = {10.1117/1.JATIS.11.2.021411},
       adsurl = {https://ui.adsabs.harvard.edu/abs/2025JATIS..11b1411S}
}

@ARTICLE{Riggs2025,
       author = {{Riggs}, A.~J. Eldorado and {Bailey}, Vanessa P. and {Moody}, Dwight and {Balasubramanian}, Kunjithapatham and {Basinger}, Scott A. and {Belikov}, Ruslan and {Bendek}, Eduardo and {Debes}, John and {Dube}, Brandon D. and {Gersh-Range}, Jessica and {Groff}, Tyler D. and {Kasdin}, N. Jeremy and {Mennesson}, Bertrand and {Monacelli}, Brian and {Moore}, Douglas M. and {Ruane}, Garreth and {Sandhu}, Jagmit and {Shi}, Fang and {Sidick}, Erkin and {Siegler}, Nicholas and {Sirbu}, Dan and {Trauger}, John and {Weisberg}, Carey L. and {White}, Victor E. and {Wilson}, Daniel W. and {Wilson}, Robert C. and {Yee}, Karl Y. and {Zimmerman}, Neil T.},
        title = "{Flight masks of the Roman Space Telescope Coronagraph Instrument}",
      journal = {Journal of Astronomical Telescopes, Instruments, and Systems},
         year = 2025,
        month = apr,
       volume = {11},
          eid = {021403},
        pages = {021403},
          doi = {10.1117/1.JATIS.11.2.021403},
archivePrefix = {arXiv},
       eprint = {2508.08063},
 primaryClass = {astro-ph.IM},
       adsurl = {https://ui.adsabs.harvard.edu/abs/2025JATIS..11b1403R}
}

@INPROCEEDINGS{Bailey2023,
       author = {{Bailey}, Vanessa P. and {Bendek}, Eduardo and {Monacelli}, Brian and {Baker}, Caleb and {Bedrosian}, Gasia and {Cady}, Eric and {Douglas}, Ewan S. and {Groff}, Tyler and {Hildebrandt}, Sergi R. and {Kasdin}, N. Jeremy and {Krist}, John and {Macintosh}, Bruce and {Mennesson}, Bertrand and {Morrissey}, Patrick and {Poberezhskiy}, Ilya and {Subedi}, Hari B. and {Rhodes}, Jason and {Roberge}, Aki and {Ygouf}, Marie and {Zellem}, Robert T. and {Zhao}, Feng and {Zimmerman}, Neil T.},
        title = "{Nancy Grace Roman Space Telescope coronagraph instrument overview and status}",
    booktitle = {Society of Photo-Optical Instrumentation Engineers (SPIE) Conference Series},
         year = 2023,
       series = {Society of Photo-Optical Instrumentation Engineers (SPIE) Conference Series},
       volume = {12680},
        month = oct,
          eid = {126800T},
        pages = {126800T},
          doi = {10.1117/12.2679036},
archivePrefix = {arXiv},
       eprint = {2309.08672},
 primaryClass = {astro-ph.IM},
       adsurl = {https://ui.adsabs.harvard.edu/abs/2023SPIE12680E..0TB}
}

@ARTICLE{Ruane2020,
       author = {{Ruane}, Garreth and {Wallace}, J. Kent and {Steeves}, John and {Prada}, Camilo Mejia and {Seo}, Byoung-Joon and {Bendek}, Eduardo and {Coker}, Carl and {Chen}, Pin and {Crill}, Brendan and {Jewell}, Jeff and {Kern}, Brian and {Marx}, David and {Poon}, Phillip K. and {Redding}, David and {Riggs}, A.~J. Eldorado and {Siegler}, Nicholas and {Zimmer}, Robert},
        title = "{Wavefront sensing and control in space-based coronagraph instruments using Zernike's phase-contrast method}",
      journal = {Journal of Astronomical Telescopes, Instruments, and Systems},
         year = 2020,
        month = oct,
       volume = {6},
          eid = {045005},
        pages = {045005},
          doi = {10.1117/1.JATIS.6.4.045005},
archivePrefix = {arXiv},
       eprint = {2010.10541},
 primaryClass = {astro-ph.IM},
       adsurl = {https://ui.adsabs.harvard.edu/abs/2020JATIS...6d5005R}
}

@INPROCEEDINGS{Sauvage2015,
       author = {{Sauvage}, Jean-Fran{\c{c}}ois and {Fusco}, Thierry and {Guesalaga}, Andres and {Wizinowitch}, Peter and {O'Neal}, Jared and {N'Diaye}, Mamadou and {Vigan}, Arthur and {Girard}, Julien and {Lesur}, Geoffrey and {Mouillet}, David and {Buezit}, Jean-Luc and {Kasper}, Markus and {Le Louarn}, Miska and {Mlli}, Julien and {Dohlen}, Kjetil and {Neichel}, Benoit and {Bourget}, Pierre and {Heigenauer}, Pierre and {Mawet}, Dimitri},
        title = "{Low Wind Effect, the main limitation of the SPHERE instrument}",
    booktitle = {Adaptive Optics for Extremely Large Telescopes IV (AO4ELT4)},
         year = 2015,
        month = oct,
          eid = {E9},
        pages = {E9},
       adsurl = {https://ui.adsabs.harvard.edu/abs/2015aoel.confE...9S}
}

@ARTICLE{Potier2020,
       author = {{Potier}, A. and {Galicher}, R. and {Baudoz}, P. and {Huby}, E. and {Milli}, J. and {Wahhaj}, Z. and {Boccaletti}, A. and {Vigan}, A. and {N'Diaye}, M. and {Sauvage}, J.-F.},
        title = "{Increasing the raw contrast of VLT/SPHERE with the dark hole technique. I. Simulations and validation on the internal source}",
      journal = {\aap},
         year = 2020,
        month = jun,
       volume = {638},
          eid = {A117},
        pages = {A117},
          doi = {10.1051/0004-6361/202038010},
archivePrefix = {arXiv},
       eprint = {2005.02179},
 primaryClass = {astro-ph.IM},
       adsurl = {https://ui.adsabs.harvard.edu/abs/2020A&A...638A.117P}
}

@ARTICLE{Potier2022,
       author = {{Potier}, A. and {Mazoyer}, J. and {Wahhaj}, Z. and {Baudoz}, P. and {Chauvin}, G. and {Galicher}, R. and {Ruane}, G.},
        title = "{Increasing the raw contrast of VLT/SPHERE with the dark hole technique. II. On-sky wavefront correction and coherent differential imaging}",
      journal = {\aap},
         year = 2022,
        month = sep,
       volume = {665},
          eid = {A136},
        pages = {A136},
          doi = {10.1051/0004-6361/202244185},
archivePrefix = {arXiv},
       eprint = {2208.11244},
 primaryClass = {astro-ph.IM},
       adsurl = {https://ui.adsabs.harvard.edu/abs/2022A&A...665A.136P}
}

@ARTICLE{Pueyo2009,
       author = {{Pueyo}, Laurent and {Kay}, Jason and {Kasdin}, N. Jeremy and {Groff}, Tyler and {McElwain}, Michael and {Give'on}, Amir and {Belikov}, Ruslan},
        title = "{Optimal dark hole generation via two deformable mirrors with stroke minimization}",
      journal = {\ao},
         year = 2009,
        month = nov,
       volume = {48},
       number = {32},
        pages = {6296},
          doi = {10.1364/AO.48.006296},
       adsurl = {https://ui.adsabs.harvard.edu/abs/2009ApOpt..48.6296P}
}

@ARTICLE{Gaudi2020,
       author = {{Gaudi}, B. Scott and {Seager}, Sara and {Mennesson}, Bertrand and {Kiessling}, Alina and {Warfield}, Keith and {Cahoy}, Kerri and {Clarke}, John T. and {Domagal-Goldman}, Shawn and {Feinberg}, Lee and {Guyon}, Olivier and {Kasdin}, Jeremy and {Mawet}, Dimitri and {Plavchan}, Peter and {Robinson}, Tyler and {Rogers}, Leslie and {Scowen}, Paul and {Somerville}, Rachel and {Stapelfeldt}, Karl and {Stark}, Christopher and {Stern}, Daniel and {Turnbull}, Margaret and {Amini}, Rashied and {Kuan}, Gary and {Martin}, Stefan and {Morgan}, Rhonda and {Redding}, David and {Stahl}, H. Philip and {Webb}, Ryan and {Alvarez-Salazar}, Oscar and {Arnold}, William L. and {Arya}, Manan and {Balasubramanian}, Bala and {Baysinger}, Mike and {Bell}, Ray and {Below}, Chris and {Benson}, Jonathan and {Blais}, Lindsey and {Booth}, Jeff and {Bourgeois}, Robert and {Bradford}, Case and {Brewer}, Alden and {Brooks}, Thomas and {Cady}, Eric and {Caldwell}, Mary and {Calvet}, Rob and {Carr}, Steven and {Chan}, Derek and {Cormarkovic}, Velibor and {Coste}, Keith and {Cox}, Charlie and {Danner}, Rolf and {Davis}, Jacqueline and {Dewell}, Larry and {Dorsett}, Lisa and {Dunn}, Daniel and {East}, Matthew and {Effinger}, Michael and {Eng}, Ron and {Freebury}, Greg and {Garcia}, Jay and {Gaskin}, Jonathan and {Greene}, Suzan and {Hennessy}, John and {Hilgemann}, Evan and {Hood}, Brad and {Holota}, Wolfgang and {Howe}, Scott and {Huang}, Pei and {Hull}, Tony and {Hunt}, Ron and {Hurd}, Kevin and {Johnson}, Sandra and {Kissil}, Andrew and {Knight}, Brent and {Kolenz}, Daniel and {Kraus}, Oliver and {Krist}, John and {Li}, Mary and {Lisman}, Doug and {Mandic}, Milan and {Mann}, John and {Marchen}, Luis and {Marrese-Reading}, Colleen and {McCready}, Jonathan and {McGown}, Jim and {Missun}, Jessica and {Miyaguchi}, Andrew and {Moore}, Bradley and {Nemati}, Bijan and {Nikzad}, Shouleh and {Nissen}, Joel and {Novicki}, Megan and {Perrine}, Todd and {Pineda}, Claudia and {Polanco}, Otto and {Putnam}, Dustin and {Qureshi}, Atif and {Richards}, Michael and {Eldorado Riggs}, A.~J. and {Rodgers}, Michael and {Rud}, Mike and {Saini}, Navtej and {Scalisi}, Dan and {Scharf}, Dan and {Schulz}, Kevin and {Serabyn}, Gene and {Sigrist}, Norbert and {Sikkia}, Glory and {Singleton}, Andrew and {Shaklan}, Stuart and {Smith}, Scott and {Southerd}, Bart and {Stahl}, Mark and {Steeves}, John and {Sturges}, Brian and {Sullivan}, Chris and {Tang}, Hao and {Taras}, Neil and {Tesch}, Jonathan and {Therrell}, Melissa and {Tseng}, Howard and {Valente}, Marty and {Van Buren}, David and {Villalvazo}, Juan and {Warwick}, Steve and {Webb}, David and {Westerhoff}, Thomas and {Wofford}, Rush and {Wu}, Gordon and {Woo}, Jahning and {Wood}, Milana and {Ziemer}, John and {Arney}, Giada and {Anderson}, Jay and {Ma{\'\i}z-Apell{\'a}niz}, Jes{\'u}s and {Bartlett}, James and {Belikov}, Ruslan and {Bendek}, Eduardo and {Cenko}, Brad and {Douglas}, Ewan and {Dulz}, Shannon and {Evans}, Chris and {Faramaz}, Virginie and {Feng}, Y. Katherina and {Ferguson}, Harry and {Follette}, Kate and {Ford}, Saavik and {Garc{\'\i}a}, Miriam and {Geha}, Marla and {Gelino}, Dawn and {G{\"o}tberg}, Ylva and {Hildebrandt}, Sergi and {Hu}, Renyu and {Jahnke}, Knud and {Kennedy}, Grant and {Kreidberg}, Laura and {Isella}, Andrea and {Lopez}, Eric and {Marchis}, Franck and {Macri}, Lucas and {Marley}, Mark and {Matzko}, William and {Mazoyer}, Johan and {McCandliss}, Stephan and {Meshkat}, Tiffany and {Mordasini}, Christoph and {Morris}, Patrick and {Nielsen}, Eric and {Newman}, Patrick and {Petigura}, Erik and {Postman}, Marc and {Reines}, Amy and {Roberge}, Aki and {Roederer}, Ian and {Ruane}, Garreth and {Schwieterman}, Edouard and {Sirbu}, Dan and {Spalding}, Christopher and {Teplitz}, Harry and {Tumlinson}, Jason and {Turner}, Neal and {Werk}, Jessica and {Wofford}, Aida and {Wyatt}, Mark and {Young}, Amber and {Zellem}, Rob},
        title = "{The Habitable Exoplanet Observatory (HabEx) Mission Concept Study Final Report}",
      journal = {arXiv e-prints},
         year = 2020,
        month = jan,
          eid = {arXiv:2001.06683},
        pages = {arXiv:2001.06683},
archivePrefix = {arXiv},
       eprint = {2001.06683},
 primaryClass = {astro-ph.IM},
       adsurl = {https://ui.adsabs.harvard.edu/abs/2020arXiv200106683G}
}

@misc{LUVOIR2019,
       author = {{The LUVOIR Team}},
        title = "{The LUVOIR Mission Concept Study Final Report}",
      journal = {arXiv e-prints},
         year = 2019,
        month = dec,
          eid = {arXiv:1912.06219},
        pages = {arXiv:1912.06219},
archivePrefix = {arXiv},
       eprint = {1912.06219},
 primaryClass = {astro-ph.IM},
       adsurl = {https://ui.adsabs.harvard.edu/abs/2019arXiv191206219T}
}

@ARTICLE{Soummer2007,
       author = {{Soummer}, R{\'e}mi and {Ferrari}, Andr{\'e} and {Aime}, Claude and {Jolissaint}, Laurent},
        title = "{Speckle Noise and Dynamic Range in Coronagraphic Images}",
      journal = {\apj},
         year = 2007,
        month = nov,
       volume = {669},
       number = {1},
        pages = {642-656},
          doi = {10.1086/520913},
archivePrefix = {arXiv},
       eprint = {0706.1739},
 primaryClass = {astro-ph},
       adsurl = {https://ui.adsabs.harvard.edu/abs/2007ApJ...669..642S}
}

@INPROCEEDINGS{Juanola-Parramon2019,
       author = {{Juanola-Parramon}, Roser and {Zimmerman}, Neil T. and {Pueyo}, Laurent and {Bolcar}, Matthew and {Ruane}, Garreth and {Krist}, John and {Groff}, Tyler},
        title = "{The LUVOIR Extreme Coronagraph for Living Planetary Systems (ECLIPS) II. Performance evaluation, aberration sensitivity analysis and exoplanet detection simulations}",
    booktitle = {Society of Photo-Optical Instrumentation Engineers (SPIE) Conference Series},
         year = 2019,
       series = {Society of Photo-Optical Instrumentation Engineers (SPIE) Conference Series},
       volume = {11117},
        month = sep,
          eid = {1111702},
        pages = {1111702},
          doi = {10.1117/12.2530356},
       adsurl = {https://ui.adsabs.harvard.edu/abs/2019SPIE11117E..02J}
}

@ARTICLE{Redmond2024,
       author = {{Redmond}, Susan F. and {Pueyo}, Laurent A. and {Por}, Emiel H. and {Pourcelot}, Rapha{\"e}l. and {Laginja}, Iva and {Nickson}, Bryony and {Sahoo}, Ananya and {Nguyen}, Meiji M. and {Kasdin}, N. Jeremy and {Perrin}, Marshall D. and {Soummer}, R{\'e}mi and {Pogorelyuk}, Leonid},
        title = "{Exoplanet detection techniques for direct imaging dark zone maintenance datasets}",
      journal = {Journal of Astronomical Telescopes, Instruments, and Systems},
         year = 2024,
        month = oct,
       volume = {10},
          eid = {049003},
        pages = {049003},
          doi = {10.1117/1.JATIS.10.4.049003},
       adsurl = {https://ui.adsabs.harvard.edu/abs/2024JATIS..10d9003R}
}

@INPROCEEDINGS{Krist2007,
       author = {{Krist}, John E.},
        title = "{PROPER: an optical propagation library for IDL}",
    booktitle = {Optical Modeling and Performance Predictions III},
         year = 2007,
       editor = {{Kahan}, Mark A.},
       series = {Society of Photo-Optical Instrumentation Engineers (SPIE) Conference Series},
       volume = {6675},
        month = sep,
          eid = {66750P},
        pages = {66750P},
          doi = {10.1117/12.731179},
       adsurl = {https://ui.adsabs.harvard.edu/abs/2007SPIE.6675E..0PK}
}

@ARTICLE{Chambouleyron2024,
       author = {{Chambouleyron}, Vincent and {Ciss{\'e}}, Mahawa and {Salama}, Ma{\"\i}ssa and {Haffert}, Sebastiaan and {D{\'e}o}, Vincent and {Guthery}, Charlotte and {Wallace}, J. Kent and {Dillon}, Daren and {Jensen-Clem}, Rebecca and {Hinz}, Phil and {Macintosh}, Bruce},
        title = "{Reconstruction methods for the phase-shifted Zernike wavefront sensor}",
      journal = {arXiv e-prints},
         year = 2024,
        month = sep,
          eid = {arXiv:2409.04547},
        pages = {arXiv:2409.04547},
          doi = {10.48550/arXiv.2409.04547},
archivePrefix = {arXiv},
       eprint = {2409.04547},
 primaryClass = {astro-ph.IM},
       adsurl = {https://ui.adsabs.harvard.edu/abs/2024arXiv240904547C}
}
\bibliographystyle{spiebib} 

\end{document}